\documentclass[aip,cha,amsmath,reprint,superscriptaddress]{revtex4-1}
\usepackage{graphicx}
\usepackage{latexsym}
\usepackage{amsfonts}
\usepackage{color}
\usepackage{subfigure}
\usepackage{lineno}

\begin{document}

\title{Mixed mode synchronization and network bursting of
neurons with post-inhibitory rebound}
\date{\today}

\author{Roman Nagornov}
\email{nagornov.r@gmail.com}
\affiliation{Department of Control Theory, Nizhni Novgorod State 
University, Gagarin Av. 23,
606950, Nizhni Novgorod, Russia}

\author{Grigory Osipov}
\affiliation{Department of Control Theory, Nizhni Novgorod State 
University, Gagarin Av. 23,
606950, Nizhni Novgorod, Russia}

\author{Maxim Komarov}
\affiliation{Department of Control Theory, Nizhni Novgorod State 
University, Gagarin Av. 23,
606950, Nizhni Novgorod, Russia}
\affiliation{Department of Physics and Astronomy, University of Potsdam,
Karl-Liebknecht-Str 24/25, Potsdam, Germany}

\author{Arkady Pikovsky}
\affiliation{Department of Control Theory, Nizhni Novgorod State 
University, Gagarin Av. 23,
606950, Nizhni Novgorod, Russia}
\affiliation{Department of Physics and Astronomy, University of Potsdam,
Karl-Liebknecht-Str 24/25, Potsdam, Germany}

\author{Andrey Shilnikov}
\affiliation{Department of Control Theory, Nizhni Novgorod State 
University, Gagarin Av. 23,
606950, Nizhni Novgorod, Russia}
\affiliation{Neuroscience Institute and Department of Mathematics and Statistics, Georgia State University,
100 Piedmont Str., Atlanta, GA 30303, USA}

\begin{abstract}

We study activity rhythms produced by a half-center oscillator -- a pair of reciprocally coupled neurons that are capable of producing post-inhibitory rebounds.  Network (coupling-induced) burstings possessing two time scales, one of fast spiking and another of slow bursting and quiescent periods, are shown to exhibit nontrivial synchronization properties. We consider several network configurations composed of both endogenously bursting, tonic spiking and quiescent neurons, as well as various  mixed combinations. We show that synchronization at low 
frequency of bursting can be accompanied by complex,  sometimes chaotic, patters of fast spikes. 
\end{abstract}

\maketitle


{\bf 

This study is focused on the mechanisms of rhythmogenesis and robustness of anti-phase bursting in half-center-oscillators (HCOs) consisting of two reciprocally inhibitory coupled neurons. There is a growing body of experimental evidence that a HCO is a universal building block for larger neural networks, including central pattern generators (CPGs)  controlling a variety of
locomotion behaviors in spineless animals and mammals. It remains unclear how CPGs achieve the level of robustness and stability observed in nature.  
There has been a vastly growing consensus in the community of neurophysiologists and computational researchers that some basic structural and functional elements are likely shared by CPGs of both invertebrate and vertebrate animals.   
In this study we consider several configurations of HCOs including coupled endogenous bursters, tonic spiking, and quiescent neurons, that become network bursters only when coupled by fast inhibitory synapses through the mechanism of post-inhibitory-rebound (PIR). 
The goal is better understanding the PIR mechanism as a key component for 
robust anti-phase bursting in generic HCOs.}

\section{Introduction}

Synchronization of coupled oscillators is a fundamental phenomenon in nonlinear systems that has been observed in a wide range of diverse applications~\cite{Pikovsky-Rosenblum-Kurths-03}. The mathematical concept of synchronization, first introduced and developed for periodic oscillators \cite{shilnikov2004some}, has further been generalized for other aperiodic systems, including ones with chaotic dynamics. In life sciences, 
of a keen interest is synchronization or phase locking among oscillators
 with multiple time scales. They may include  
 mixed-mode  and slow-fast relaxation-type oscillators~\cite{MR2916308}, whose synergetic interaction 
 can give rise to the onset of a variety of synchronization 
 patterns~\cite{Jalil-Belykh-Shilnikov-10,Omelchenko-Rosenblum-Pikovsky-10,Jalil-Belykh-Shilnikov-12,Wojcik-Schwabedal-Clewley-Shilnikov-14}. 
In neuroscience, a plethora of rhythmic motor behaviors with diverse time scales, such as heartbeat, 
respiration,
chewing, and locomotion on land and in water  are produced and governed by neural networks called 
Central Pattern Generators (CPGs) \cite{CPG,Marder-Calabrese-96}. A CPG is a microcircuit of neurons 
that is able to autonomously generate an array of polyrhythmic bursting patterns, underlying various 
motor behaviors. 

Endogenous (self-sustained) bursting and network (coupling-induced) bursting  are composite 
oscillatory behaviors, featuring  active phases during which a neuron or a group of neurons 
generates trains of fast action potentials, which are alternated with long interburst intervals 
during which it remains inactive or quiescent, until a new cycle of bursting occurs.     
In this paper we examine synchronization of bursting patterns emerging through interactions of two 
interneurons coupled reciprocally by fast inhibitory synapses. This study has been driven by two 
major motivations: first, a 
general one concerning questions on synchronization of mixed-mode oscillators. The second is a 
neuroscience related one, aimed at understanding of intrinsic mechanisms of rhythmogenesis in CPGs 
composed, often symmetrically, of such small networks of interneurons,  as outlined below. It is 
still unclear how CPGs can achieve the level of synergy and robustness to produce  
a plethora of rhythmic patterns observed in nature. 

Recent experimental and theoretical studies have disclosed a distinct role of CPGs in generation of 
adaptive and coordinated motor activity of animals~\cite{Marder-Calabrese-96,Gillner-Wallen-85,Rabinovich-Abarbanel-06,Shaw-Thomas-14}.
An important feature of CPGs is their ability to produce various types of rhythmic bursting activity, 
what causes flexible and adaptive locomotion of an organism. To robustly govern motor patterns, CPGs 
are able to flexibly adjust their oscillatory properties (such as bursts  duration, frequency of 
spiking, phase relations of bursts)   due to feedback from sensory inputs, for example,  in response 
to changes of environment \cite{Marder-Calabrese-96,Gillner-Wallen-85}. Up to a certain extent, the 
flexibility of CPG behaviors may be attributed to its multistability (of several coexistent 
attractors representing  different bursting rhythms in a phase space of the 
dynamical system)  allowing 
for fast switching between operating modes~\cite{Schwabedal-Neiman-Shilnikov-14,Wojcik-Schwabedal-Clewley-Shilnikov-14}. 

From the theoretical point of view, a CPG is modeled as a small
network of coupled oscillatory, or quiescent, interneurons, each described by a system of nonlinear 
ordinary differential or difference equations (dynamical system)~\cite{Rabinovich-Abarbanel-06,Izhikevich-07}. The study of CPGs allows
one to develop a general understanding of synchronization patterns in mixed-mode
oscillators, applicable to systems of different physical and biological origin.

There is a growing body of experimental evidence that a universal building block of most identified 
CPGs is a half-center oscillator (HCO) \cite{Hill-VanHooser-Calabrese-03}. A HCO is a pair of two 
reciprocally inhibitory interneurons bursting in alternation. Such a pair can be composed of 
endogenously bursting interneurons, as well as of endogenously tonic spiking or quiescent ones that 
start bursting only when they are coupled in a network. Theoretical studies~\cite{Wang-Rinzel-92,Daun-Rubin-Rybak-09} have indicated that formation of an
anti-phase bursting rhythm is always 
based on some slow subsystem dynamics. The slow subsystem is strongly associated with the slow 
membrane current, such as persistent sodium current or slow calcium-dependent current; following \cite{DESTEXHE-CONTRERAS-SEJNOWSKI-STERIADE-94} we term currents associated with slow-varying 
concentrations and gating variables as slow ones. There are three basic mechanisms to generate 
alternating bursting in the HCO: release, escape, and post-inhibitory rebound (PIR). The first 
mechanism is typical for endogenously bursting neurons \cite{Jalil-Belykh-Shilnikov-10,Jalil-Belykh-Shilnikov-12}. The other two mechanisms underlie 
network bursting in HCOs  comprised of neurons which are depolarized or  hyperpolarized quiescent 
interneurons, respectively, in   isolation \cite{Perkel-Mulloney-74,Skinner-Kopell-Marder-94, Angstadt-Grassmann-Theriault-Levasseur-05, Marder-Calabrese-96, Kopell-Ermentrout-02} 

\begin{figure}[!ht]
\centering
\includegraphics[width=\columnwidth]{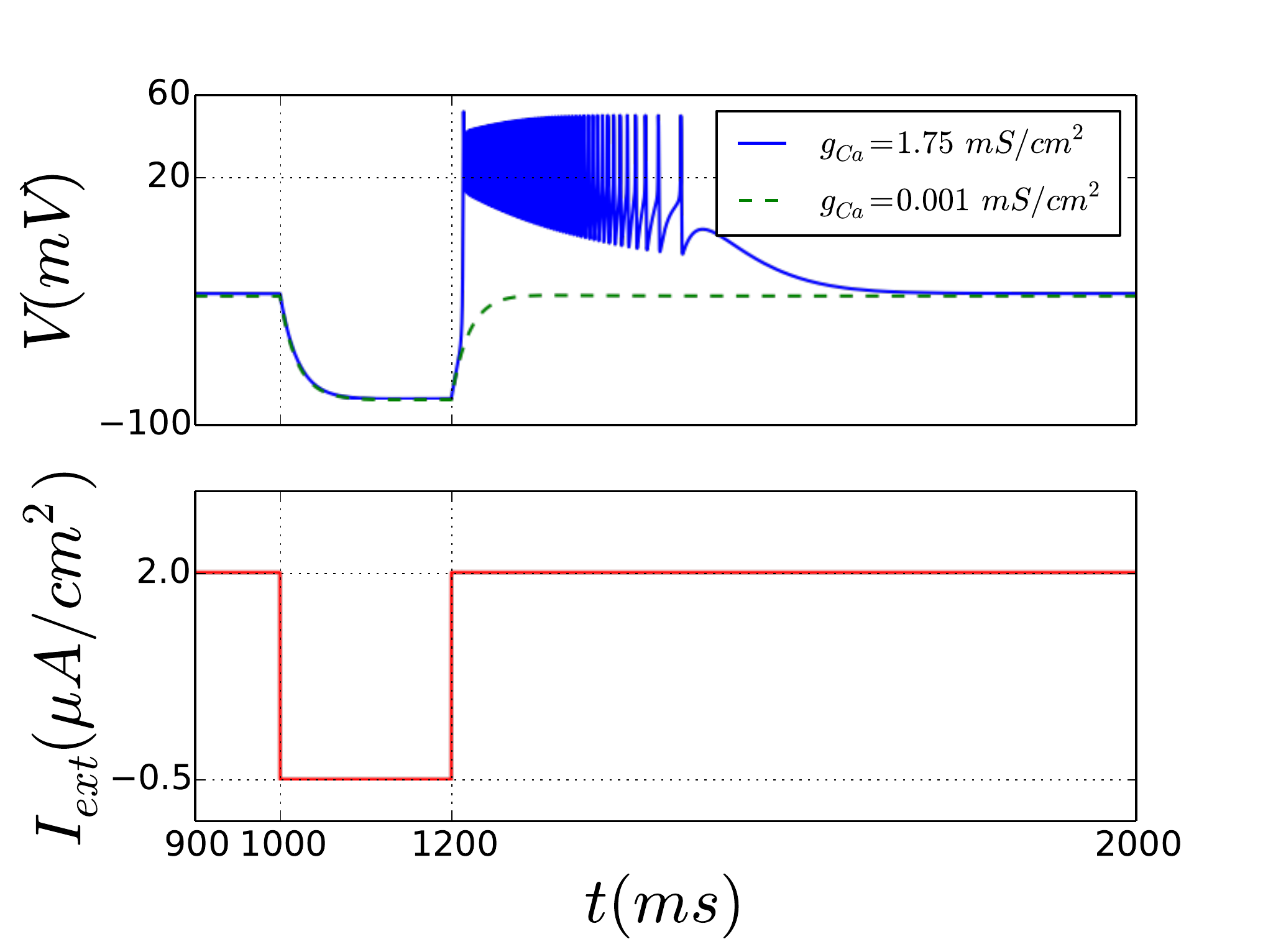}
\caption{Abrupt release of a hyperpolarized pulse of the external current $ I_{ext}$ (bottom panel) triggering a post-inhibitory rebound of bursting in the quiescent neuron (top panel). The parameter $g_{Ca}$ stated in the legend denotes maximal conductance of the current $I_T$. When the parameter $g_{Ca}$ is relatively small, the effect of the $I_T$ current is negligible what causes absence of PIR in the
dynamics of the neuron (dashed green curve in the top panel). }
\label{fig:rebound}
\end{figure}

The PIR mechanism uses reciprocal inhibition to keep network bursting. It triggers an onset of a 
single or a series of action potentials in the post-synaptic neuron after it has been prolongedly 
hyperpolarized and abruptly released from inhibition generated by the pre-synaptic neuron during an 
active, tonic spiking phase of bursting. After that, the neurons of the HCO swap their opposite roles 
to repeat the PIR mechanism. The PIR promotes the action potential generation after a period of 
sufficiently strong  hyper-polarizing (inhibiting) input, as illustrated in Fig.~\ref{fig:rebound}. 
PIR is often caused by a low-threshold activated calcium current in  neurons and their biophysically 
plausible models.   

This paper examine the PIR contribution to formation, synchronization and robustness of bursting 
rhythms  in  
inhibitory coupled neurons.  Here, we employ a modification of the Hodgkin-Huxley-type model 
introduced 
in~\cite{DESTEXHE-CONTRERAS-SEJNOWSKI-STERIADE-94}, to plausibly describe the PIR mechanism. 
Depending on its parameters, the model  is known to produce an array of typical  neuronal activities 
such as excitable dynamics, periodic spiking and bursting. 

Below we first examine the conditions that confirm and reproduce the PIR mechanism in the neurons. We 
will argue that the PIR is a pivotal component that promotes an alternating bursting rhythm in the 
HCO made of intrinsically spiking and excitable neurons.  Finally, we will show that the PIR 
mechanism makes anti-phase network bursting even more robust and easier to initiate in the case of 
endogenously bursting neurons.

The paper is organized as follows: in Section \ref{sec:model} we introduce the neuronal model, and 
discuss its dynamical properties in Section~\ref{sec:one_neuron_dynamic}. Section \ref{sync_of_2_neurons} is focused on synchronization properties of  a pair of endogenous bursters, 
while Section \ref{sync_of_2_neurons} examines anti-phase bursting onset in coupled tonic-spiking 
neurons. Sections~\ref{sec:2en} and \ref{sec:2dm} discusses the PIR-mechanisms of the HCO, and 
synergetic dynamics of mixed neurons: individually tonic spiking and hyperpolarized quiescent.  

\section{Basic Model}\label{sec:model}

In this study we use a neuronal model of the Hodgkin-Huxley type  that has been introduced in \cite{DESTEXHE-CONTRERAS-SEJNOWSKI-STERIADE-94}.
The basic equations describing the dynamics of the membrane potentials of the neurons read as follows (prime denotes the time derivative, $i=1,2$):
\begin{equation} \label{eq:beqs}
\begin{aligned}
C_{m} V^\prime_{i} &= I_{ext}^{(i)} - f_{i}(V_{i}) - \displaystyle\sum\limits_{j=1,j\neq i}^N I_{syn}(V_{i}, V_{j}),\\
f_{i}(V_{i}) &= I_{leak}^{(i)} + I_{Na}^{(i)} + I_{K}^{(i)} + I_{K[Ca]}^{(i)} + I_{T}^{(i)}. 
\end{aligned}
\end{equation}
The variable $V_i(t)$ describes evolution of the membrane potential of the $i$-th neuron. The first two terms on the right-hand side of Eq.~(\ref{eq:beqs})  govern intrinsic dynamics of the neuron: $I^{(i)}_{ext}$ stands for a constant external current applied to the neuron, while term $f_i(V_i)$ represents the sum of intrinsic ionic currents. 
The sum of synaptic currents $\sum\limits_{j=1,j\neq i}^N I_{syn}(V_{i}, V_{j})$ describes the coupling interactions between the neurons. The full details of current representation are given in the Appendix below. 

There are two principal control parameters in this system: the external current, $I_{ext}^{(i)}$,  governs the activity type in an isolated, uncoupled neuron. The second parameter, the maximal conductance $G$ of the synaptic current given by Eq.~\eqref{eq:i_syn_curr_app}, controls the 
coupling strength, and thus the network dynamics of the coupled neurons. 

The low-threshold activated calcium current $I_{T}$, modeled by Eq.~\eqref{eq:i_t} and regulated by the maximal conductance $g_{Ca} $, causes the PIR mechanism in the neurons~\eqref{eq:beqs}, this
effect is illustrated in Fig.~\ref{fig:rebound}. 
Rebound bursts can be triggered by injection of a hyper-polarizing pulse of external current $I^{(i)}_{ext}$ (bottom panel in Fig.~\ref{fig:rebound}). 
If the impact of $I_T$ current is strong enough, the neuron  generates bursts of spikes, after recovering from the hyperpolarized quiescent state  (top panel in Fig.~\ref{fig:rebound}).

\section{Dynamics of an isolated neuron}\label{sec:one_neuron_dynamic}

\begin{figure}[!ht]
  \centering
    \includegraphics[width=\columnwidth]{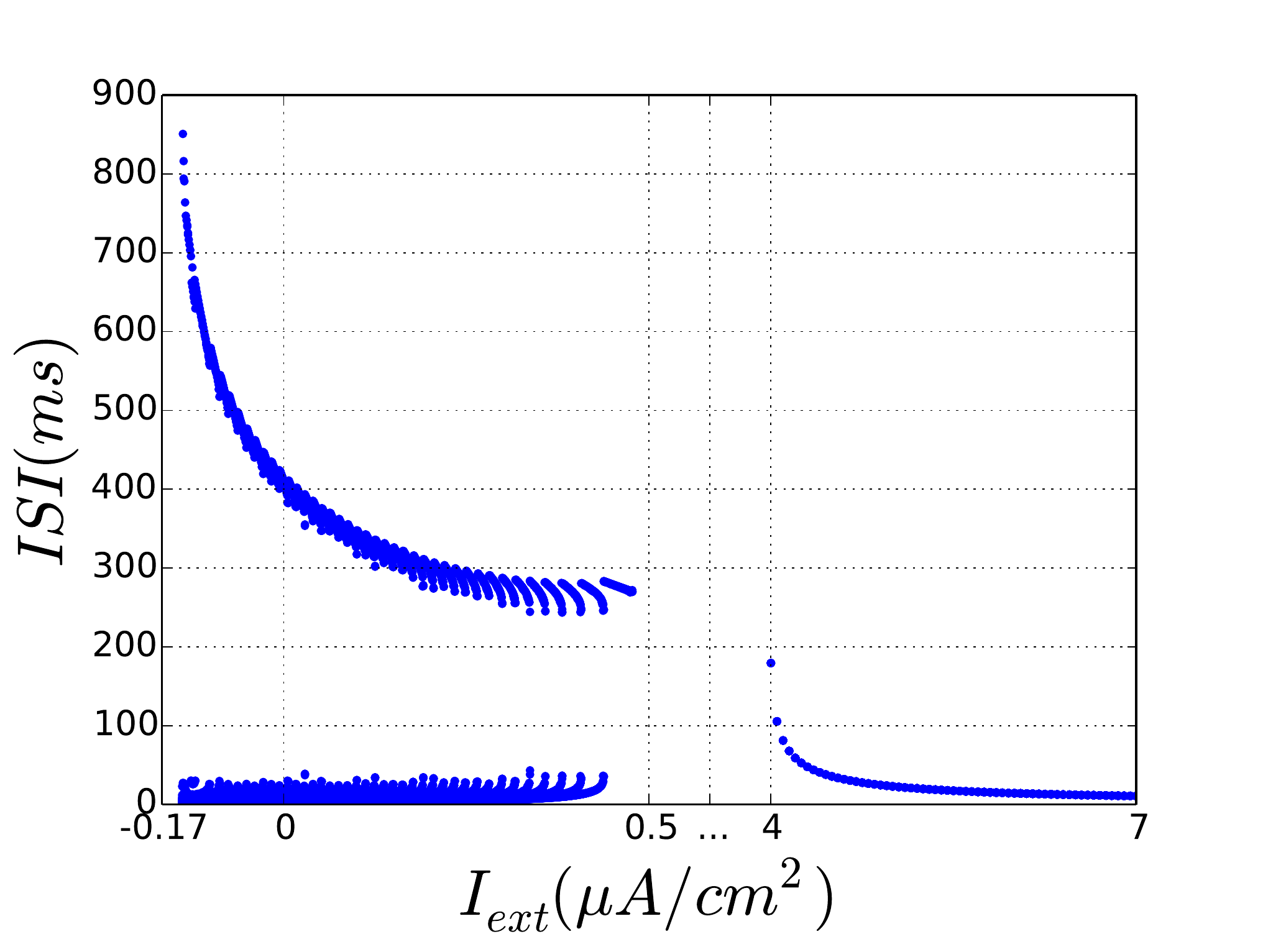}
  \caption{Bifurcation diagram for an isolated neuron: plotting inter-spike intervals (ISI) 
  against the external current $ I_{ext}$ (this horizontal axis is broken in the middle to highlight 
the ranges of nontrivial behaviors) reveals windows of bursting $(-0.14\,\mu A/ cm^{2}\lesssim 
I_{ext} 	\lesssim 0.5\,\mu A/cm^{2})$, quiescence $(0.5\,\mu A /cm^{2}\lesssim I_{ext} 	\lesssim 
4\,\mu A /cm^{2})$,  and tonic spiking $(I_{ext}\gtrsim 4\,\mu A/ cm^{2})$ activity in the neuron. 
The low ISI branch corresponds to short time intervals between fast spikes, and the top branch 
representс long interburst intervals between consecutive spike trains.}
  \label{fig:IsolatedNeuronRegimes}
\end{figure}

Let us first consider the  dynamics of an isolated neuron at $ I_{syn}=0 $ in Eq.~\eqref{eq:beqs}. We treat the external current $ I_{ext} $ as the primary bifurcation parameter in both computational and biological experiments, which allows one to examine and plausibly model various neuron activities and transitions between them. For the same purpose, we employ 
interspike intervals (ISIs), which are distances between adjacent spikes generated by the neuron, as measurements  characterizing the neuron's dynamical regimes. Figure~\ref{fig:IsolatedNeuronRegimes} serves as an illustration of how the ISIs changes with variations of $I_{ext} $.

First we describe the neuron dynamics at large  $ I_{ext}$ values. For $ I_{ext} > 4\,\mu A /cm^{2} $ 
the neuron produces tonic spiking activity (corresponding to a stable periodic orbit in the phase 
space of the model \eqref{eq:beqs}).
 Here, the value of ISI (which is the period of the stable orbit) is inversely proportional to the 
 $I_{ext}$-value. The period
decreases with increase of $ I_{ext}$, and vice-versa. As one can see from 
Fig.~\ref{fig:IsolatedNeuronRegimes}, near a critical value $ I_{ext}\approx 4\,\mu A /cm^{2} $ the 
ISI shows an ``unbounded'' growth, which is an indication 
of a bifurcation of the stable periodic orbit with an arbitrarily long period.  Detailed examinations indicate the occurrence of  
a homoclinic saddle-node bifurcation \cite{Shilnikov-Shilnikov-Turaev-Chuai-01} underlying the 
transition from tonic spiking activity to hyperpolarized quiescence, represented by a stable 
equilibrium state at low values of the membrane potential $V$. This stable equilibrium state (a node 
with real and negative characteristic exponents) persists within the parameter window   
$ 0.8\,\mu A /cm^{2} \lesssim I_{ext} \lesssim 4\,\mu A /cm^{2}$, where the neuron remains ready for PIRs.  Small perturbations of the quiescent state of the neuron 
have no pronounced effects. Relatively  strong  perturbations can trigger a spike, after which the neuron comes back to the over-damped quiescence state. 
With decreases of $ I_{ext} $ from the threshold $I_{ext}\approx 4\,\mu A /cm^{2}$, the steady state becomes a focus.  The stable focus loses the stability through a supercritical Andronov-Hopf bifurcation at $I_{ext} \approx 0.8\,\mu A /cm^{2}$. 
These oscillations are not seen in Fig.~\ref{fig:IsolatedNeuronRegimes} because they are under the 
threshold of spike generation.  The amplitude of the oscillations increases rapidly as $ I_{ext} $ is 
further decreased. At $ I_{ext} \approx 0.5\,\mu A /cm^{2}$ the shape of the periodic orbit in the 
phase space changes via the addition of  ``turns, '' the so-called spike-adding mechanism \cite{Channell2007a,Channell2009}.  Such a periodic orbit  is associated with bursting
activity, with turns corresponding to spikes within a burst.  
Bursting that includes two time scales can be recognized in Fig.~\ref{fig:IsolatedNeuronRegimes}  
through two characteristic branches: the bottom branch corresponds to small ISI values due to fast spiking, while the top branch is  due to long interburst intervals. More details on geometry of bursting can be found in ~\cite{Izhikevich-07,Shilnikov-12}.
Decreasing the external current $ I_{ext}$ makes the bursting neuron more depolarized  and increases the number of spikes per burst. 
Having reached some maximal value of spikes per burst at $ I_{ext} \approx -0.136\,\mu A /cm^{2} $, decreasing  $I_{ext}$ gives rise to bursts with fewer spikes.  
The value $ I_{ext} \approx -0.14\,\mu A /cm^{2} $ corresponds to the occurrence of another saddle-node bifurcation of equilibria since the time interval between two consecutive bursts becomes arbitrarily large. For $ I_{ext}<  -0.14\,\mu A /cm^{2} $ the neuron remains quiescent and  excitable. 

The examination of bifurcations  have been repeated  for several values of the maximal conductance, $ g_{Ca}$, regulating the low-threshold $ Ca^{2+}$ current responsible for PIR in the neuron. We have found that this current bears  an insignificant affect on the intrinsic dynamics of the individual neurons, mainly because the membrane potential does not decrease below the threshold value to activate the current. As we will see below, this is not the case for coupled neurons, where the parameter $g_{Ca}$ has a pronounced effect on the collective dynamics.

\section{Synchronization of two bursting neurons}\label{sync_of_2_neurons}

In this section and  sections~\ref{sec:2tn}-\ref{sec:2dm} below, we explore a repertoire of  rhythmic bursting outcomes generated by a HCO of 
two neurons, coupled reciprocally by fast, non-delayed inhibitory synapses.  It is known that such coupled,  endogenous 
bursters can produce a range of synchronous rhythmic outcomes with various fixed phase-lags due to overlapping spike interactions,  see   \cite{Jalil-Belykh-Shilnikov-12,jalil2013} and references therein. 

We will first examine the cooperative dynamics in the network of two coupled neurons ($N=2$ in Eqs.~(\ref{eq:beqs})) with different  $I^{(1,2)}_{ext}$  within the interval $[ -0.14\,\mu A /cm^{2},\, 0.8\,\mu A /cm^{2}]$. This range of $I_{ext}$  corresponds to endogenous bursting in both neurons.  Due to the difference,  $\Delta I_{ext}= I_{ext}^{(2)} - I_{ext}^{(1)}$, 
the temporal and quantitative  characteristics of endogenous bursters such as their period, duty cycles, the spike numbers per burst, are different. The strength of coupling is quantified by the maximal conductance, $G$, of the inhibitory synaptic current $ I_{syn}$. In what follows, we show that while $\Delta I_{ext}$ remains relatively small, increasing the coupling strength shall give rise to an onset of synchrony between the neurons with the same bursting period. However, this will no longer be true for larger $\Delta I_{ext}$  values.  

\subsection{Nearly identical bursters}\label{small mismatch}

\begin{figure}[!ht]
  \centering
    \includegraphics[width=\columnwidth]{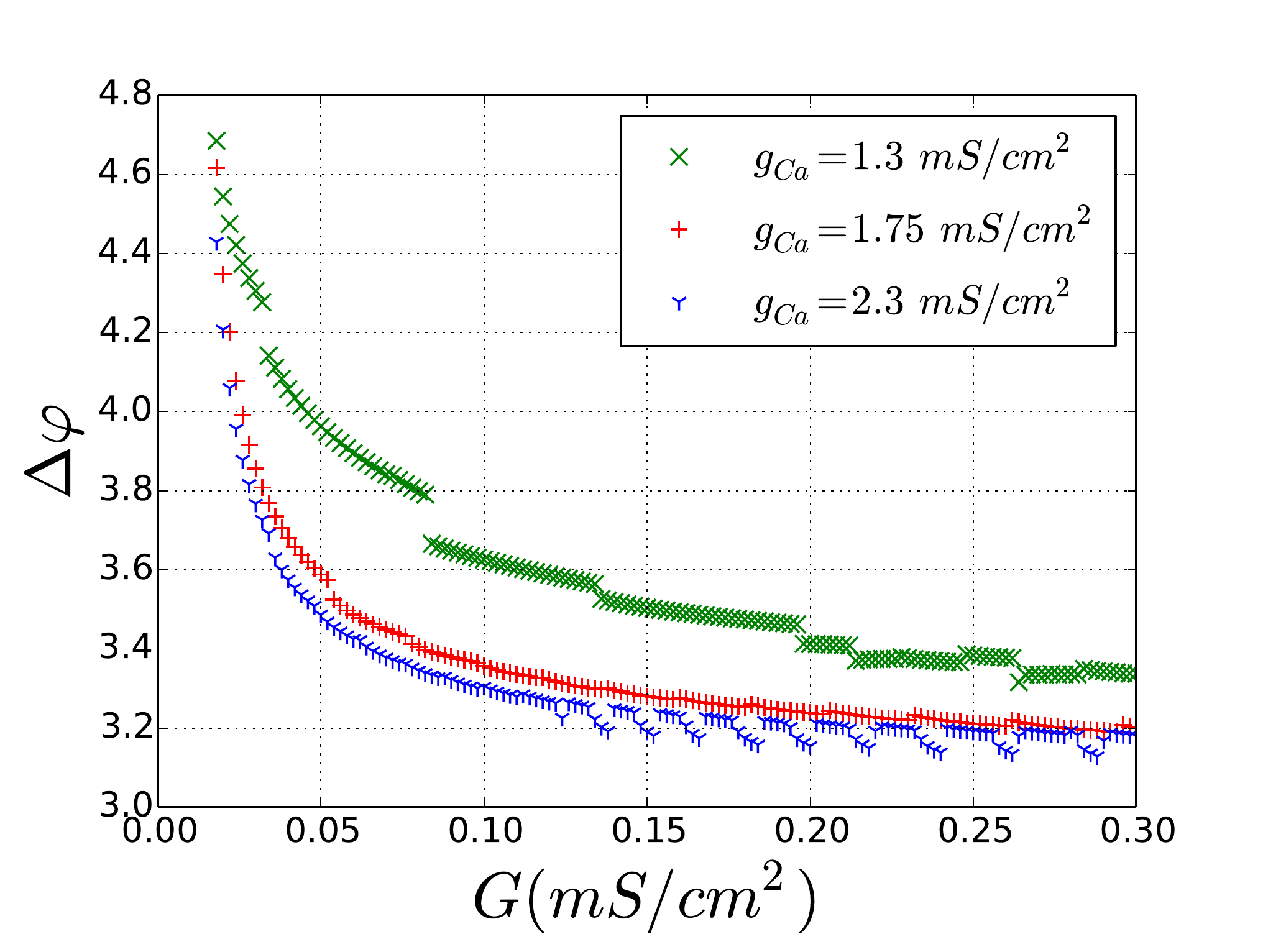}
  \caption{Mean phase lag, $\Delta \varphi$ plotted against the coupling strength  $G$ for different values of the maximal conductance $ g_{Ca} $ of the low-threshold $ Ca^{2+} $-current (larger $g_{Ca}$ values promote stronger PIRs in the neurons); here  $I_{ext}^{(1)} = 0.2\,\mu A /cm^{2}$ and  $I_{ext}^{(2)} = 0.15\,\mu A /cm^{2} $.}
  \label{fig:fig_dp}
\end{figure}

While $|\Delta I_{ext}|\lesssim 0.05\,\mu A /cm^{2}$,  synchronous bursting can  already occur at a relatively weak coupling. Here, the burst periods of both neurons become equal, however the spike number per burst in  the neurons  may not be the same. This is shown  in Fig.~\ref{fig:fig_dp}, which depicts the dependence of the mean phase-lag between oscillations of the low-threshold 
$Ca^{2+}$-currents in the  neurons, on the coupling strength $G$ for several values of the conductance $g_{Ca}$. 
Note that the post-inhibitory rebounds in the neurons under consideration are due to the slowest, low-threshold $ Ca^{2+} $-current $I_{T}$, whose magnitude is  controlled by the maximal conductance $g_{Ca}$.  It is worth noticing that PIRs occur  more reliably with increasing $g_{Ca}$.  

\begin{figure}[!ht]
\centerline{
\includegraphics[width=\columnwidth]{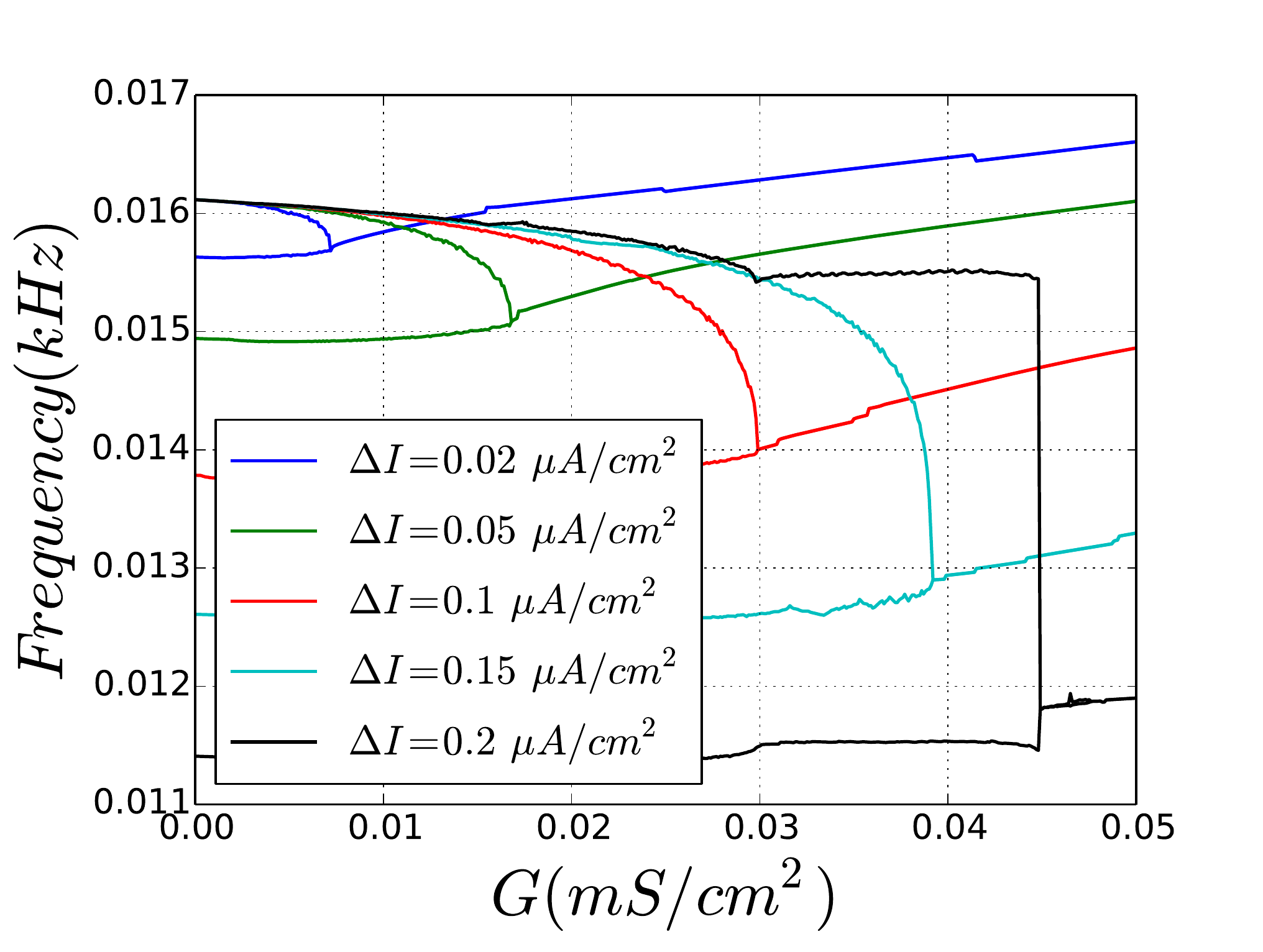}}
\caption{Bifurcation diagram for antiphase synchronization regimes: averaged frequency of bursting is plotted against the coupling strength for several increasing $\Delta I_{ext}$ values. Double overlapping branches are the indication of bursting  dichotomy with two slow different frequencies in the two coupled neurons. Note a large plateau of the pronounced 4:3 frequency locking at $ \Delta I_{ext} = 0.2\,\mu A /cm^{2}$, collapsing into anti-phase synchrony locked at an 1:1 ratio at higher $G$ values.}
	\label{fig:transition}
\end{figure}

Nevertheless, for a broad range of initial condition, the neurons will predominantly burst in anti-
phase rather than in-phase. This implies that the HCO can be bi-stable. This observation may shed  
light on properties of multifunctional CPGs composed of coupled neurons including coupled HCOs, which 
can switch 
between multiple functional states associated with different locomotion types. 
While in-phase bursting appears to be atypical for neurons with fast inhibitory coupling like in our 
case, the 
coexistence  of  anti-phase and in-phase regimes have been reported elsewhere \cite{Jalil-Belykh-Shilnikov-10,Jalil-Belykh-Shilnikov-12}. However, in-phase synchrony can prevail whenever both  HCO 
neurons are driven externally by another inhibiting burster or HCO~\cite{prl08,Shilnikov2008b}.   

Next, we discuss the effects that the PIR activity  of the neurons may have on antiphase bursting. 
Figure~\ref{fig:fig_dp} discloses how the coupling threshold for the onset of synchronous bursting depends on mismatches between the neurons of the HCO, as well as on the phase-locked frequency of slow $I_{T}$-oscillations. The increasing phases, defined on modulo $2\pi$, of the neurons are reset at the very beginning of each burst. Figure~\ref{fig:fig_dp} shows the evolution of the phase lag, $\Delta \varphi$, between the oscillations by the slow gating variables, $h_{T}^{(1,2)}$, for inactivation of the low-threshold $Ca^{2+}$-current, which is is plotted against the coupling constant $G$ for various  
$ g_{Ca}$-values.  We found in most cases that the phase lag tends eventually to $\pi$, i.e. the anti-phase bursting occurs in the HCO, with the coupling strength above a critical value. Depending on $I_{ext}$, the actual $\Delta \varphi$-value may deviate from $\pi$. One can see from Fig.~\ref{fig:fig_dp} that increasing $ g_{Ca} $ (promoting stronger  post-inhibitory rebound activity) leads to quite symmetric antiphase bursting even at small values of the coupling strength. This is an  explicit manifestation of the contribution of the slow  low-threshold $ Ca^{2+} $-current in fostering the antiphase synchronization between the bursting neurons in the HCO.

\begin{figure}[!ht]
\centering
\includegraphics[width=0.45\columnwidth]{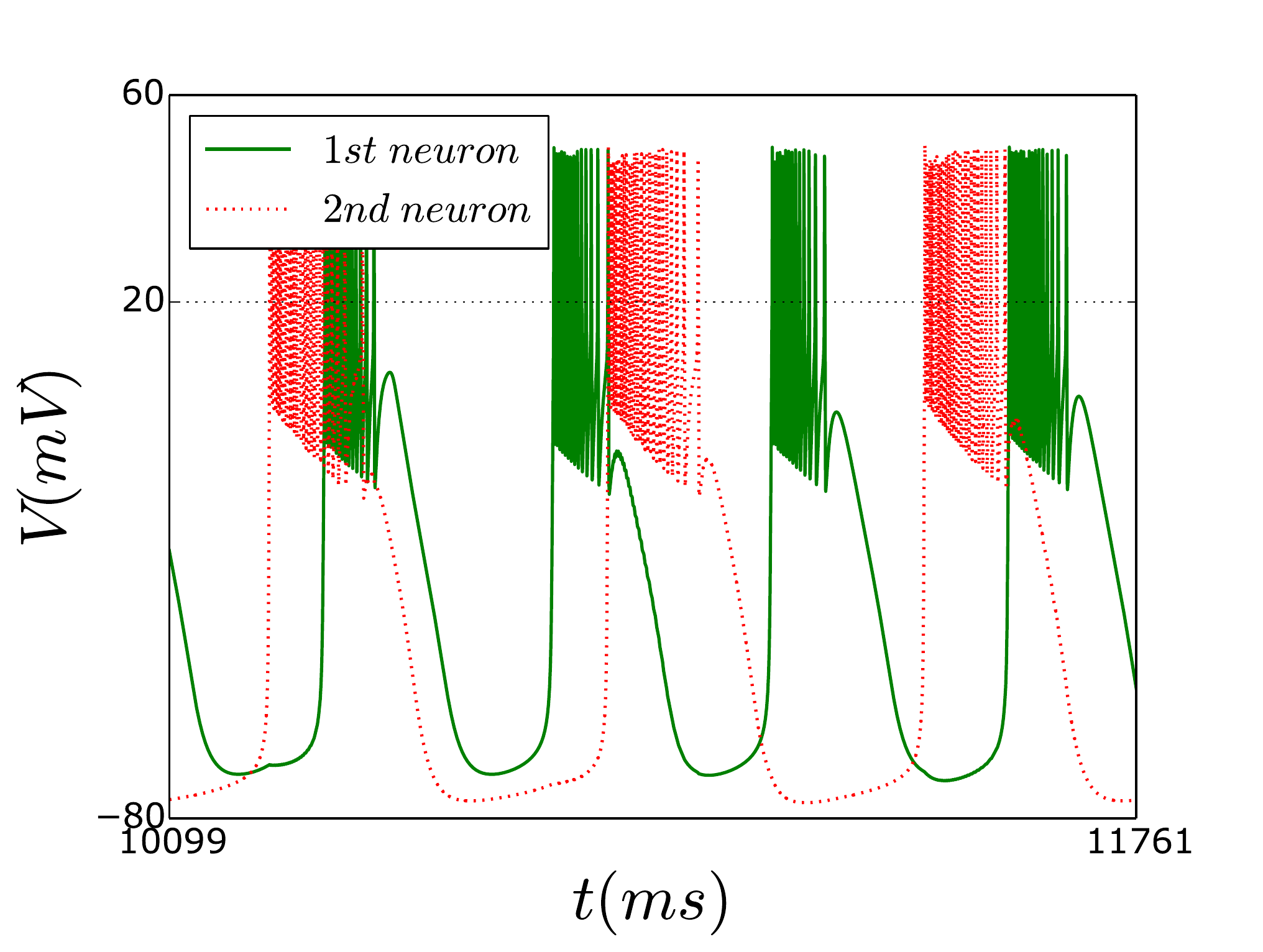}\qquad
\includegraphics[width=0.45\columnwidth]{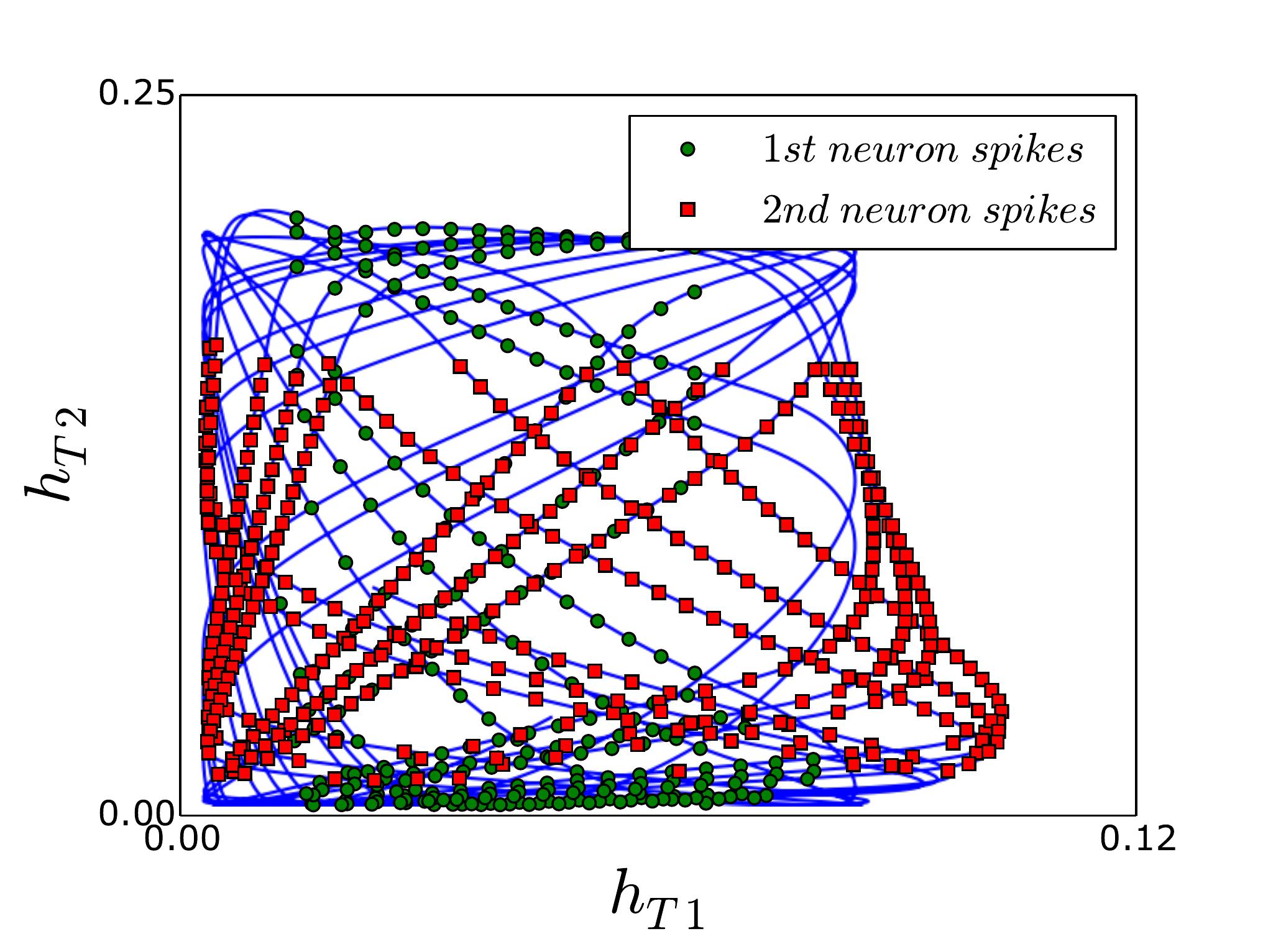}\\
\includegraphics[width=0.45\columnwidth]{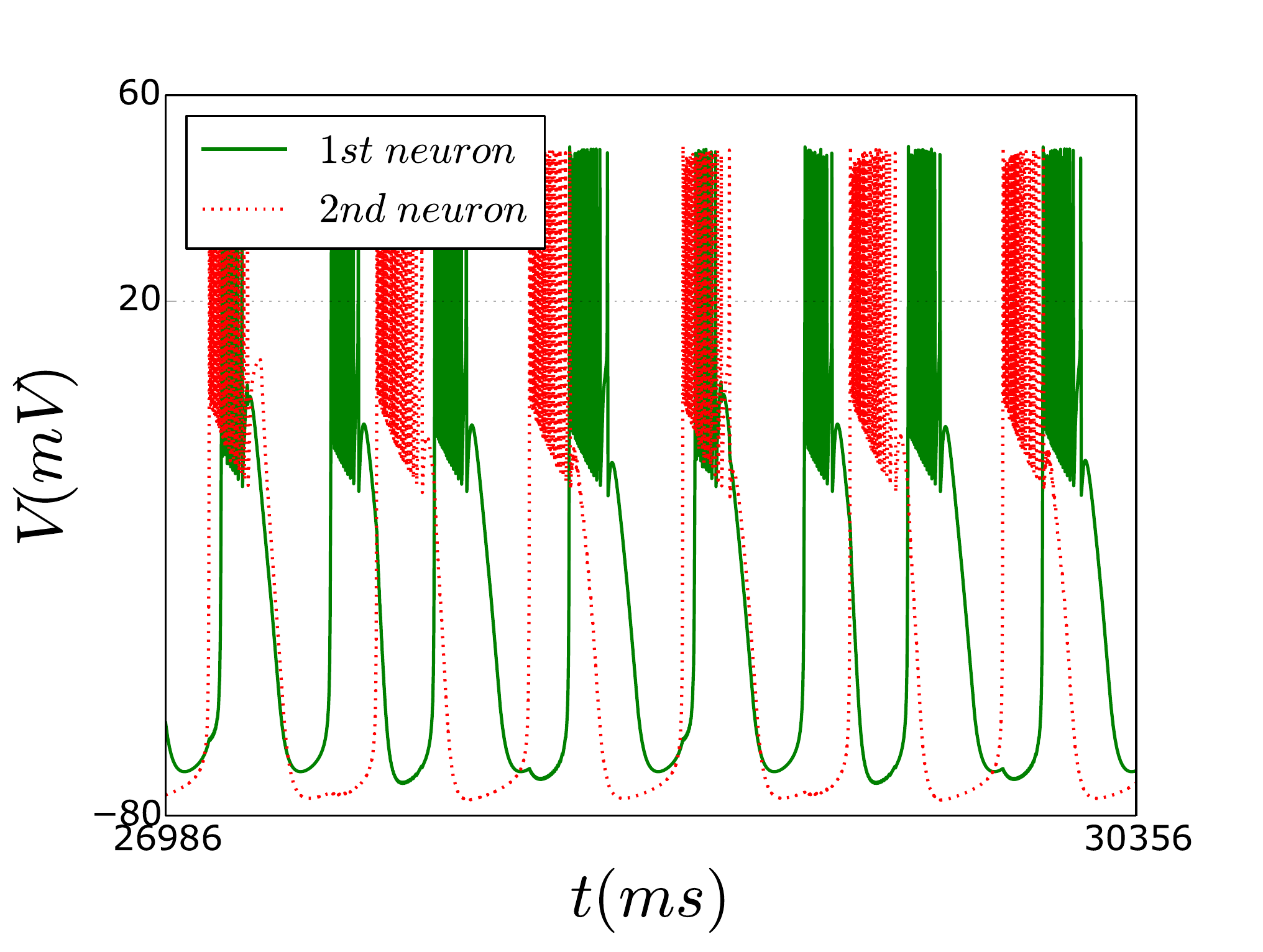}\qquad
\includegraphics[width=0.45\columnwidth]{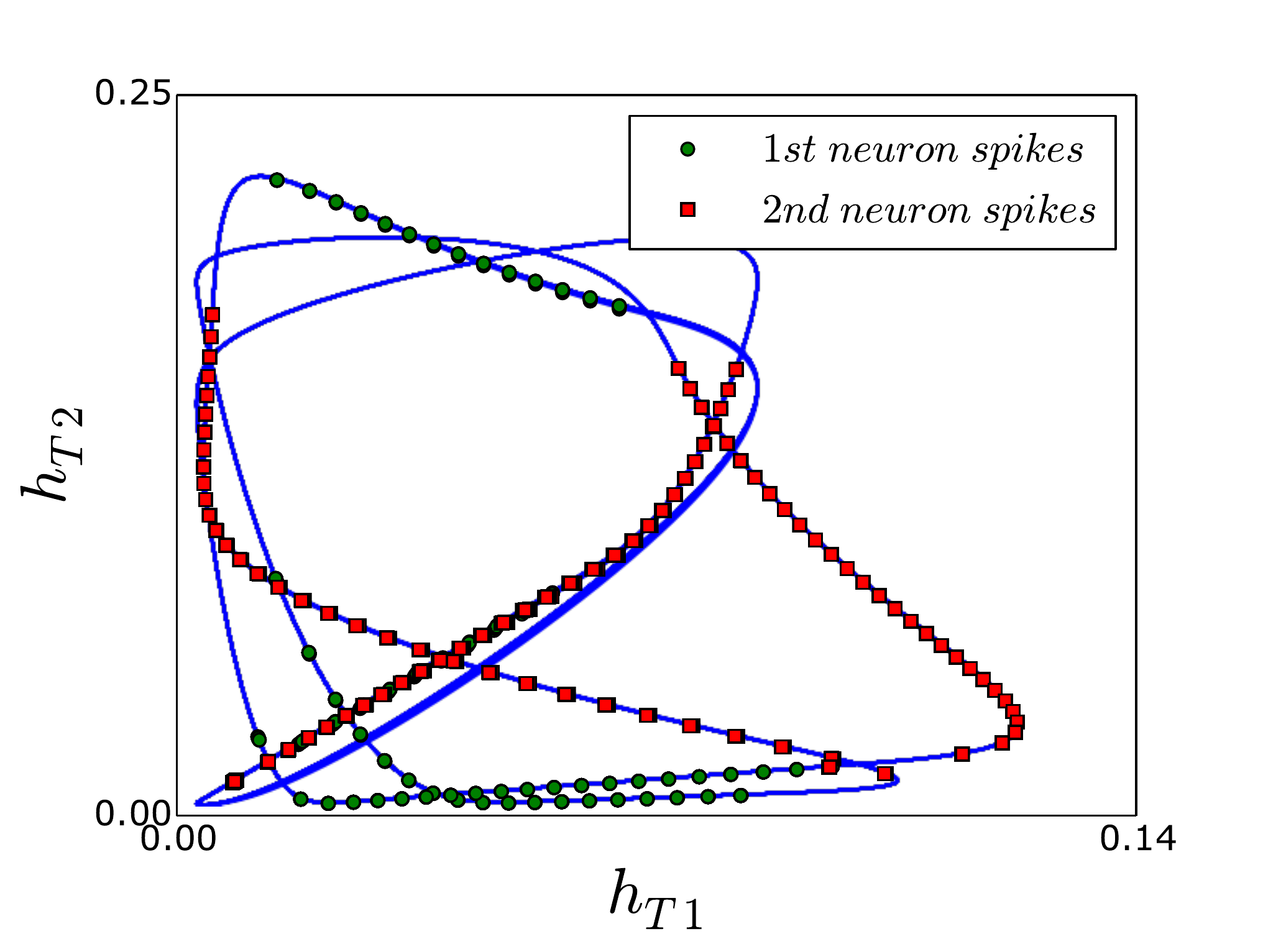}\\
\includegraphics[width=0.45\columnwidth]{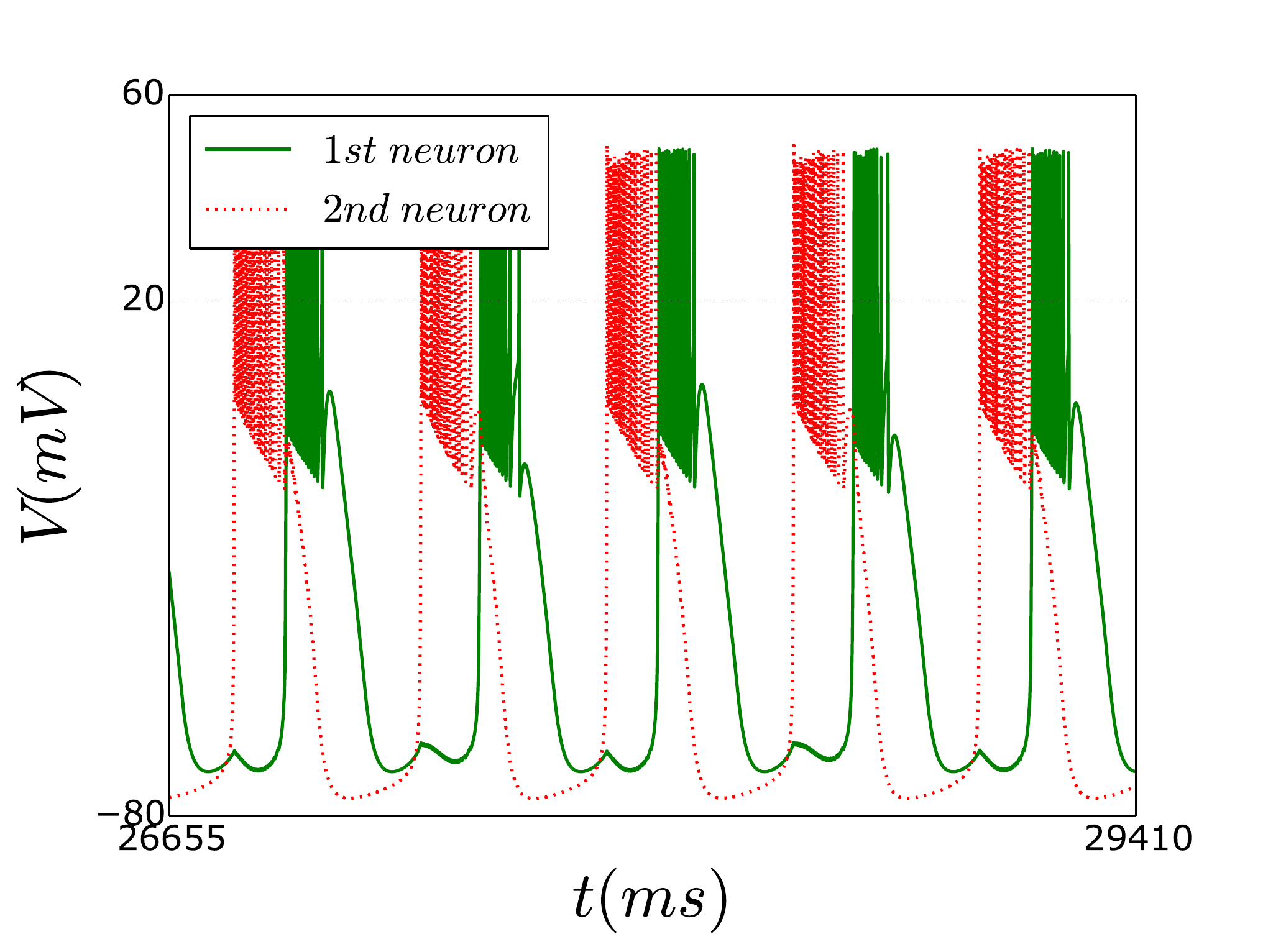}\qquad
\includegraphics[width=0.45\columnwidth]{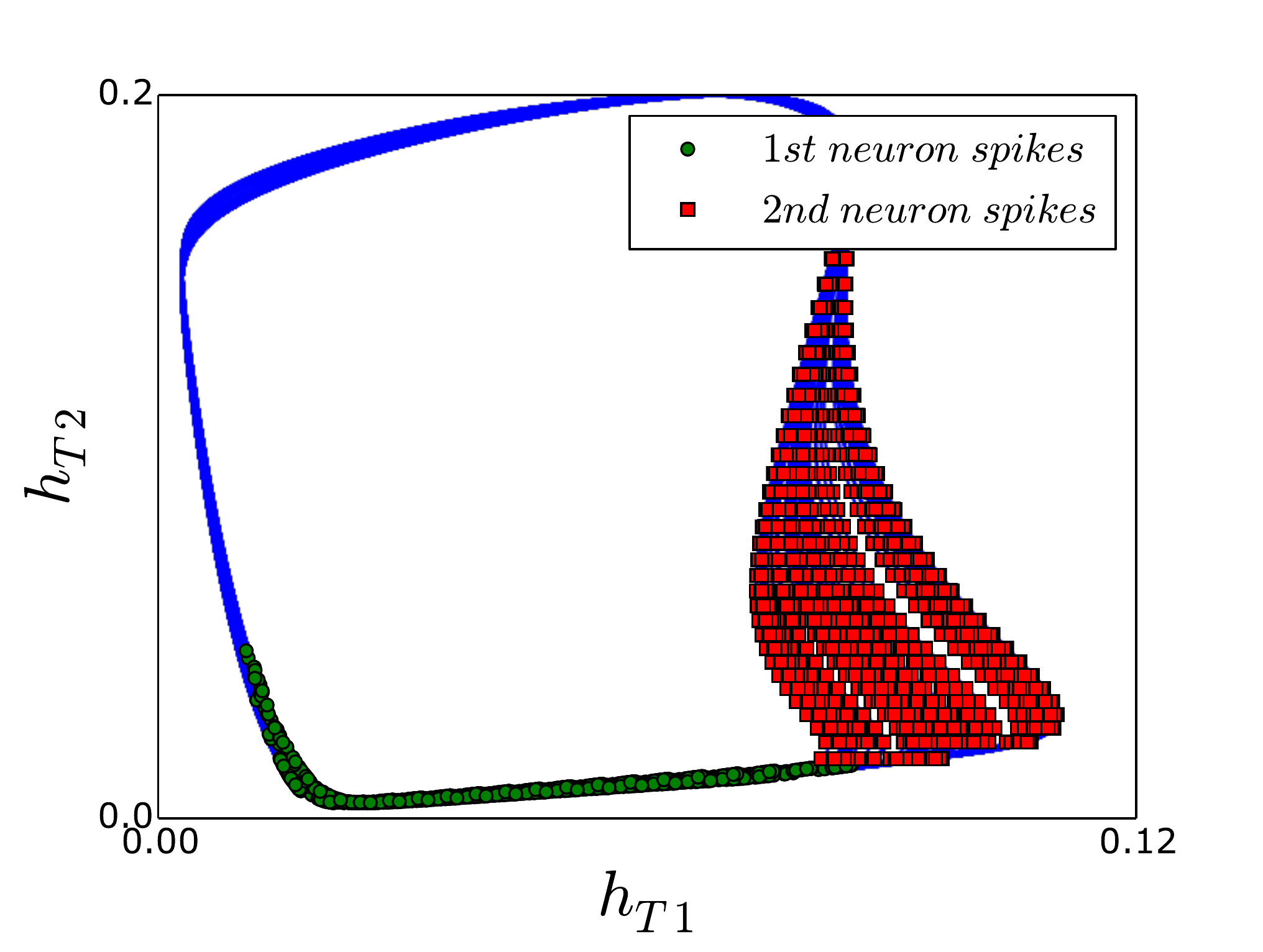}\\
\includegraphics[width=0.45\columnwidth]{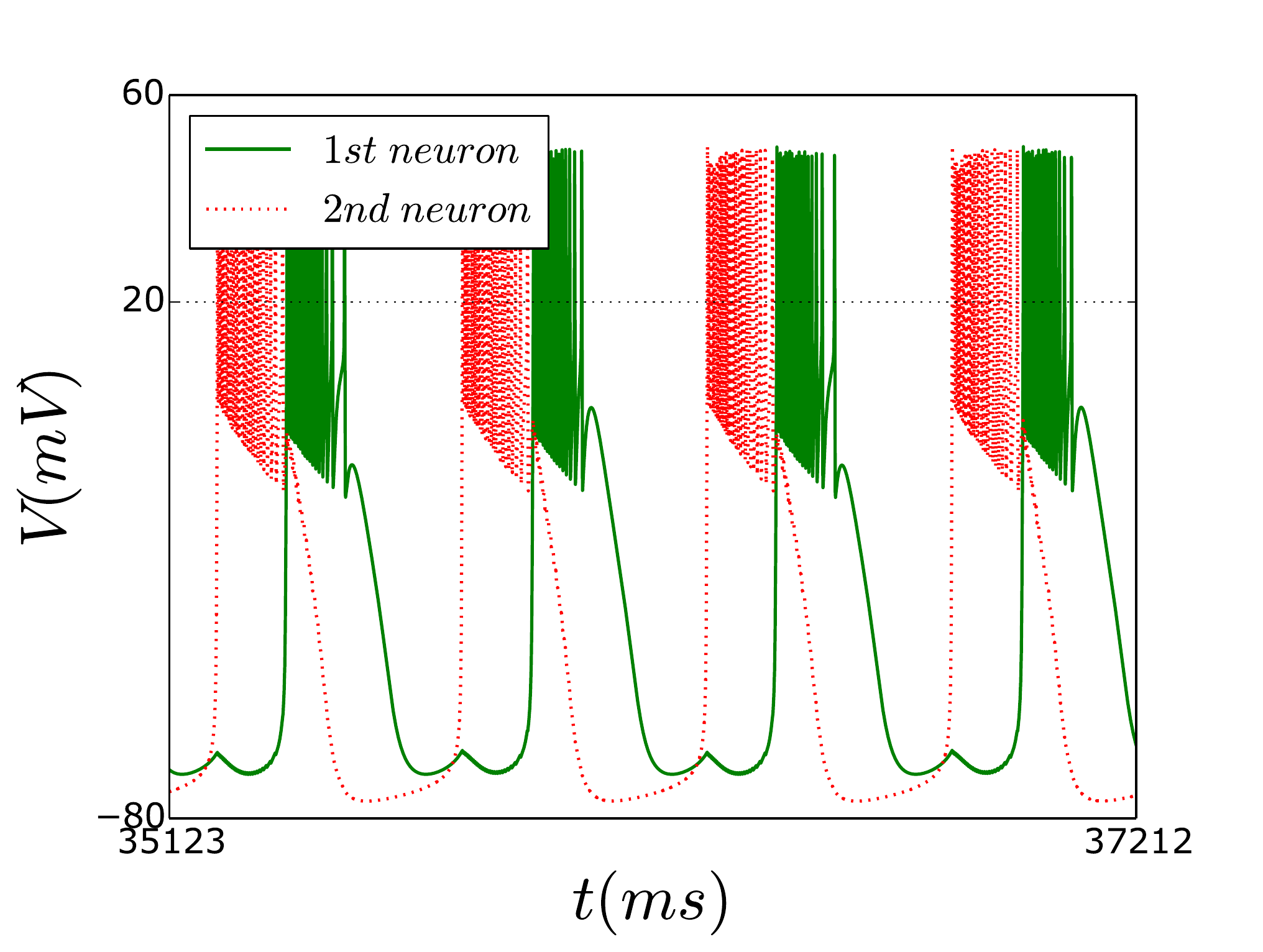}\qquad
\includegraphics[width=0.45\columnwidth]{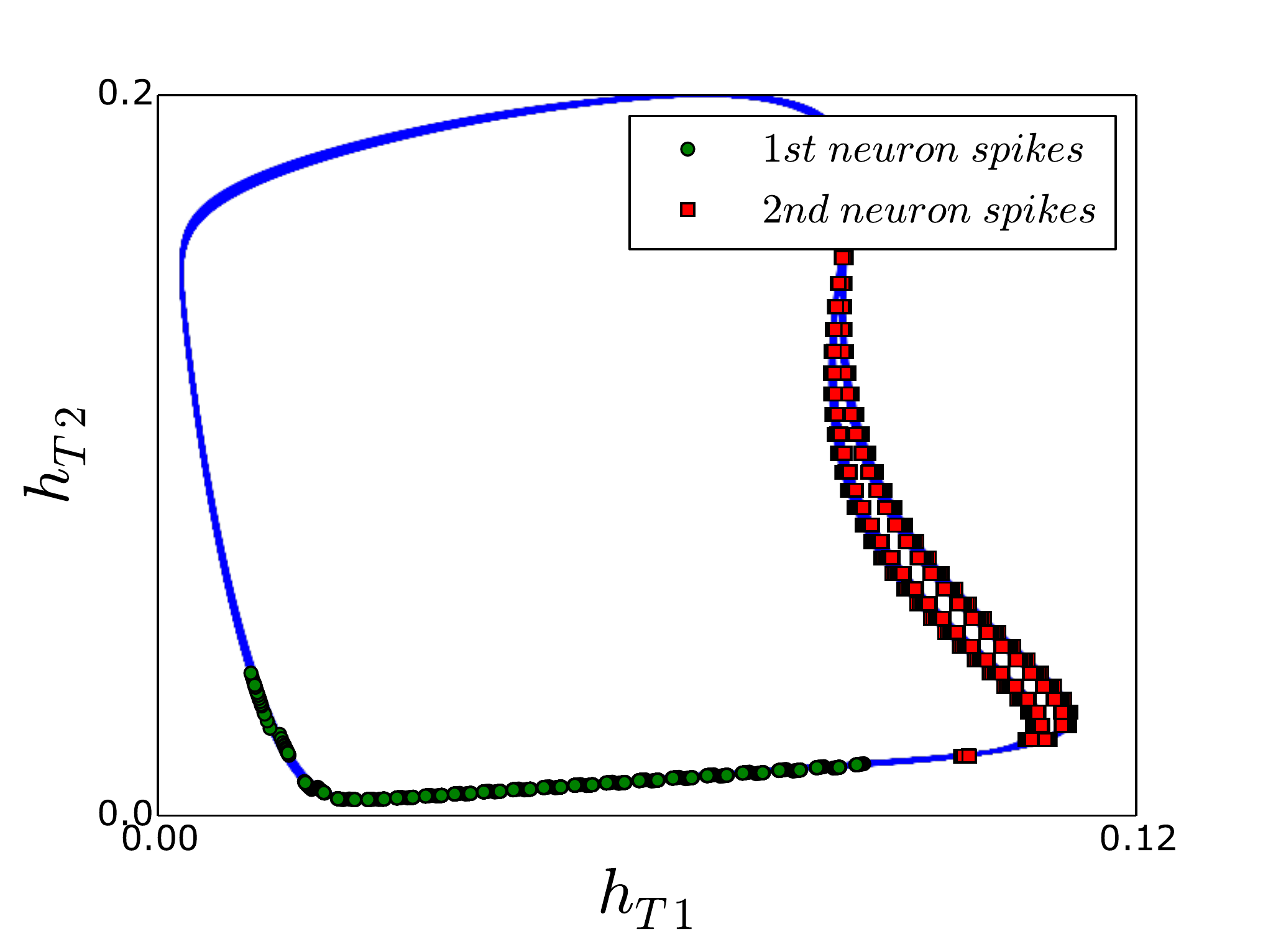}
\caption{Left column shows progressions of phase locked voltage traces of two coupled neurons at $\Delta I_{ext}=0.2\,\mu A /cm^{2} $, $ I_{ext}^{(1)} = 0.2\,\mu A /cm^{2}$
for $ G = 0.02\,mS/cm^{2};\,0.0445\,mS/cm^{2}; \,0.464\,mS/cm^{2};\, 0.048\,mS/cm^{2}$. Right panels demonstrate the Lissajous curves drawn by slow variables $h^{(1)}_T$ 
and $h^{(2)}_T$, which correspond to (top) quasi-periodic dynamics  at $ G = 0.02 $; a 4:3-frequency locking regime at $G=0.0445\,mS/cm^{2}$; 
an 1:1 chaotic locking 
at $G=0.464\,mS/cm^{2}$; (bottom) an 1:1 periodic locking  at $G=0.048\,mS/cm^{2}$. Circles and squares mark spike events in bursting.}
	\label{fig:hysteresis_allG}
\end{figure}

\subsection{Bursters far from identical}\label{large mismatch}

Increasing $\Delta I_{ext}$ between coupled oscillators leads to an array of pronounced synchronization effects. 
Figure~\ref{fig:transition} represents the slow (bursting) frequencies plotted against coupling strength, $G$, at 
several $\Delta I_{ext}$ values. One observes the presence of overlapping branches at low $G$-values. This indicates that the coupled neurons generate bursting activities at different frequencies, until the coupling is increased over a threshold value at which 
both branches merge. This threshold value  becomes higher with increases of $\Delta I_{ext}$, as the individual neurons become more and more different.    We note from this figure that for the largest case $ \Delta I_{ext}=0.2\,\mu A /cm^{2}$, the HCO neurons become locked at the 4:3 frequency ratio, prior to the occurrence of the ultimate 1:1 frequency locking at higher coupling values.

Figure~\ref{fig:hysteresis_allG} illustrates transitions to synchrony for the largest case $ \Delta I_{ext}=0.2\,\mu A /cm^{2}$.  The left panels demonstrate the established antiphase bursting in voltage traces  produced by the HCO at various values of the coupling strength, $G$, below and above the threshold. The right panels represent the so-called Lissajous curves, which are parametrically traced down by the slow variables $h^{(1)}_T$ and $h^{(2)}_T$ of both neurons.  These curves help to interpret the corresponding types of frequency locking including quasi-periodic dynamics.  On these curves we mark timing of individual spikes.

\begin{figure}[!ht]
\centering
\includegraphics[width=0.8\columnwidth]{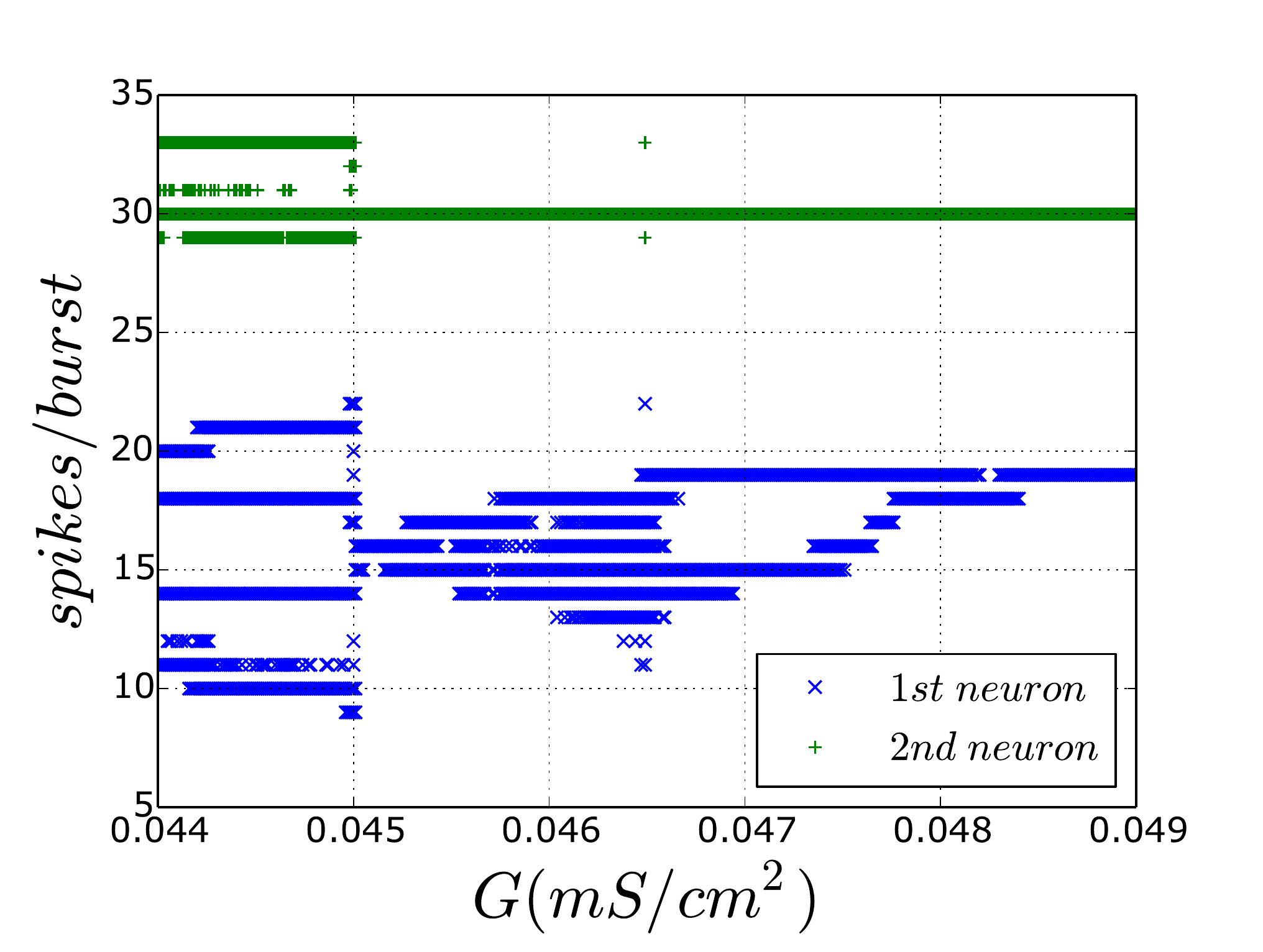}
\caption{Bifurcation diagram showing spike number per burst generated by the neurons plotted against the coupling constant $G$; other parameters are same as  in Fig.~\ref{fig:hysteresis_allG}. Value $G\approx 0.045\,mS/cm^{2}$ is a threshold towards the 1:1-frequency locking; $G\approx 0.0464\,mS/cm^{2}$
 corresponds to phase slipping in  the	chaotic 1:1 locking state of alternating bursting.}
\label{fig:LargeMismatch}
\end{figure}

It is worth noticing that such a large $ \Delta I_{ext}$  makes the phases of slow oscillations shift closer to $\pi/2$, rather than to 
$\pi$. With a $\pi/2$ phase shift, active phases of bursting in the neurons partially overlap, showing the emergence of complex dynamics caused by spike interactions. To understand these dynamics, we evaluated the distribution of numbers pf  spikes per burst for several coupling strengths. Our findings  are presented in Fig.~\ref{fig:LargeMismatch}. It is clear from this figure that the transition to the 1:1 locking occurs at $G\approx 0.045mS/cm^{2}$. Moreover, our simulation indicated that there is a small window, $\Delta G\approx 1.5\times 10^{-4}mS/cm^{2}$, of hysteresis occurring at the transition. Figure~\ref{fig:LargeMismatch} reveals a pecular feature of the synchronous state, existing at strong coupling  $G \in [ 0.045\,mS/cm^{2}, \, 0.0485\,mS/cm^{2}]$: while 
the spike number in bursts generated by neuron 2 remains nearly constant,  the number of spikes in bursts by neuron 2 deviates quite strongly before it becomes fixed with stronger coupling. 

From the two bottom insets in Fig.~\ref{fig:hysteresis_allG} one can observe  that depending on coupling strengths the 1:1 locking  features chaotic and periodic modulations via the slow gating variables of the neurons. In the chaotic case,  the number of spikes per burst in either neuron or both neurons will vary while the alternating bursting in the HCO persist locked at an 1:1-ratio within a windows $[0.046\,mS/cm^{2}, \, 0.047\,mS/cm^{2}]$. This is an explicit manifestation of the phenomena of chaotic phase synchronization~\cite{Omelchenko-Rosenblum-Pikovsky-10}. 

\section{Antiphase bursting out of spiking neurons}\label{sec:2tn}
\begin{figure}[!ht]
\centering
\subfigure [][] {\resizebox*{.45\columnwidth}{!}{\includegraphics{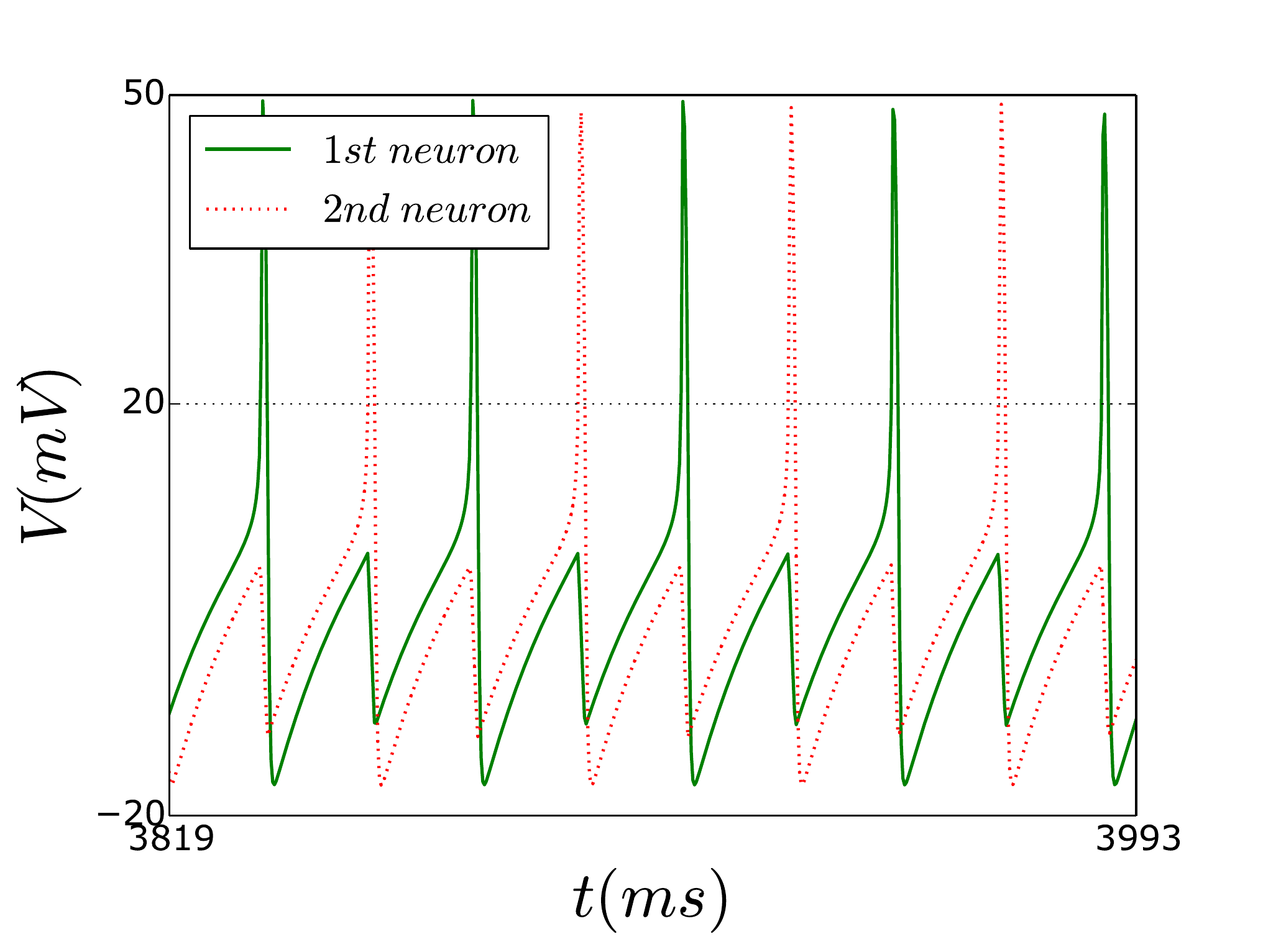}}} 
\subfigure [][] {\resizebox*{.45\columnwidth}{!}{\includegraphics{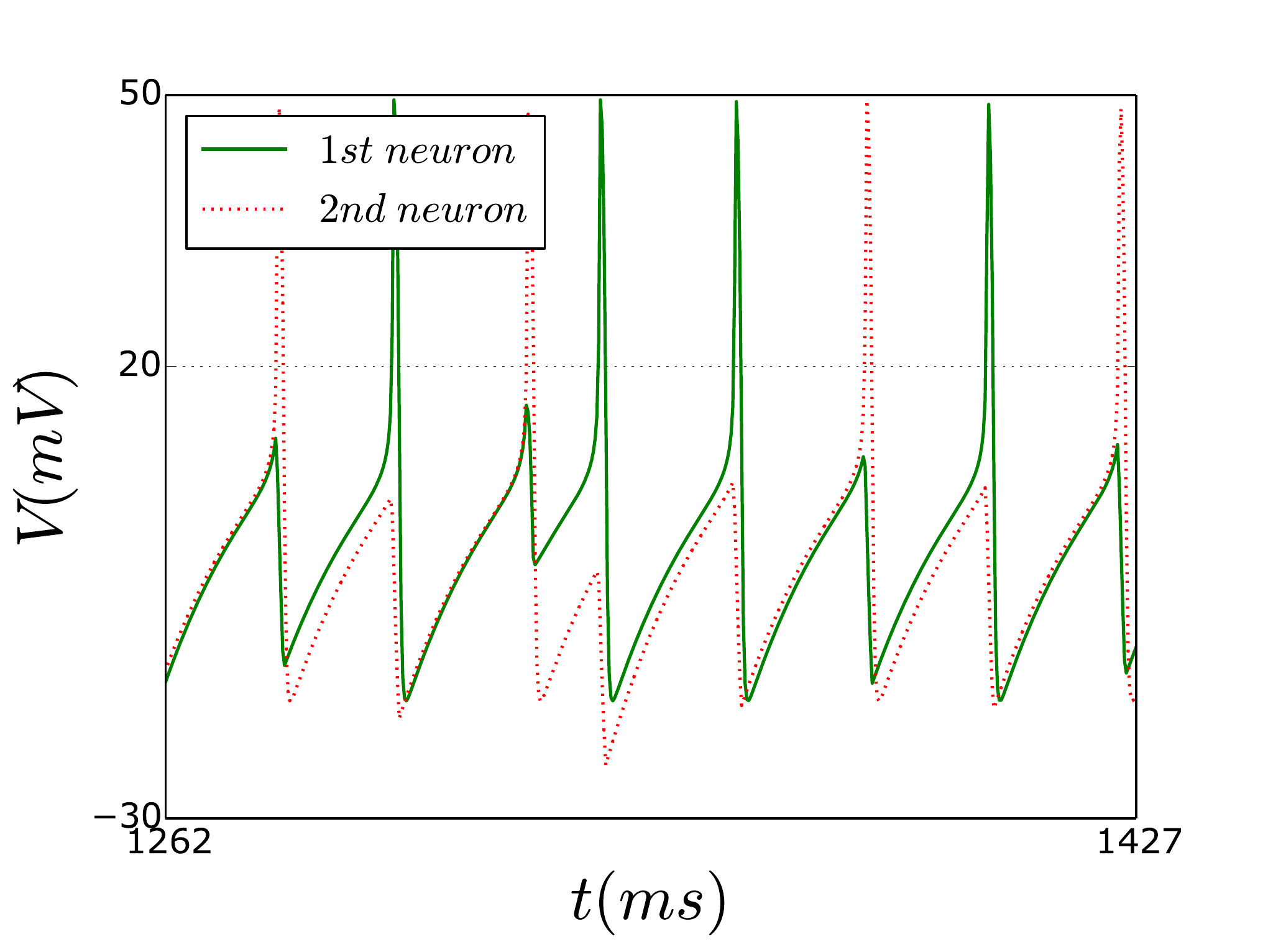}}} 
\subfigure [][] {\resizebox*{.45\columnwidth}{!}{\includegraphics{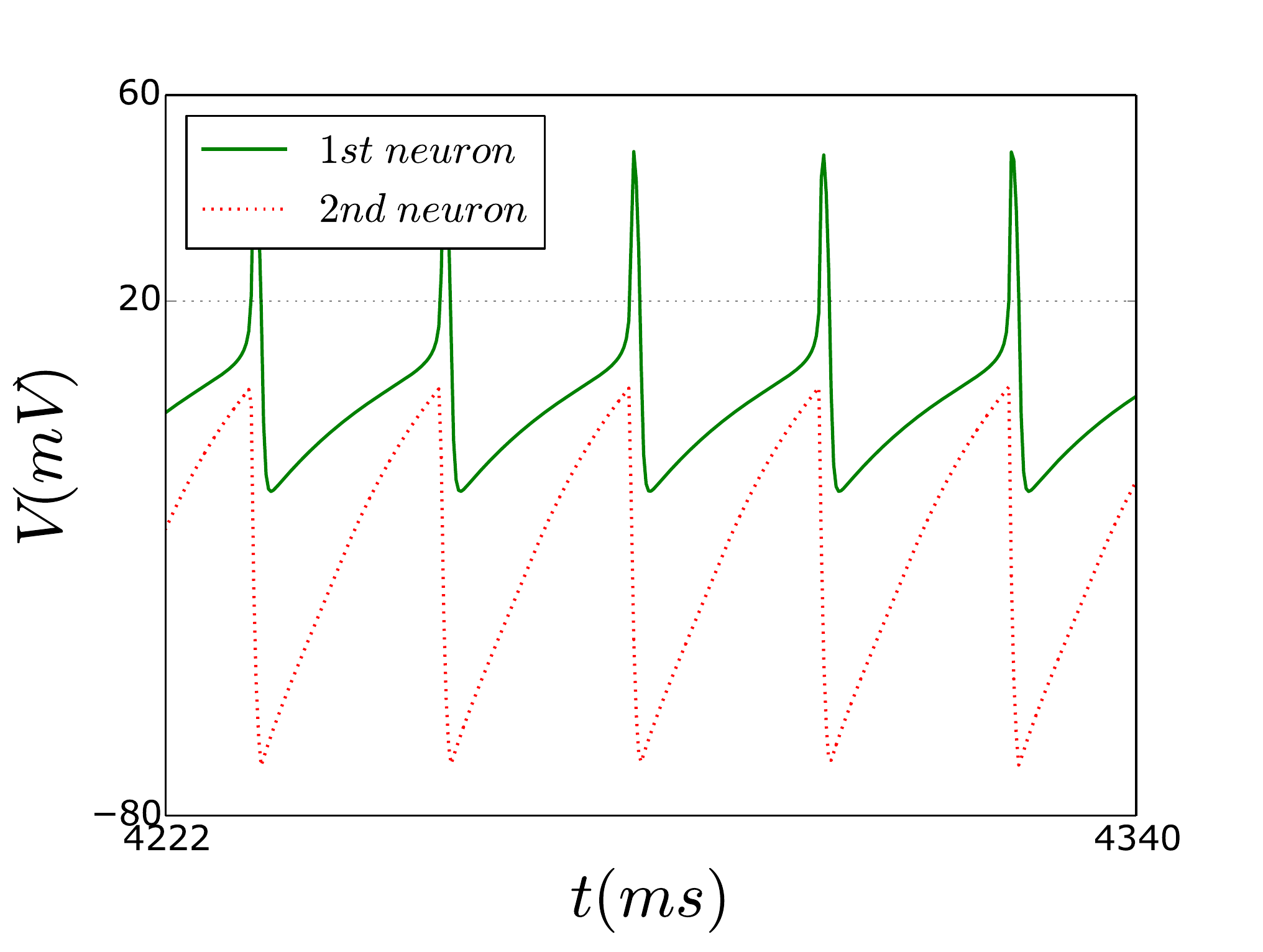}}} 
\subfigure [][] {\resizebox*{.45\columnwidth}{!}{\includegraphics{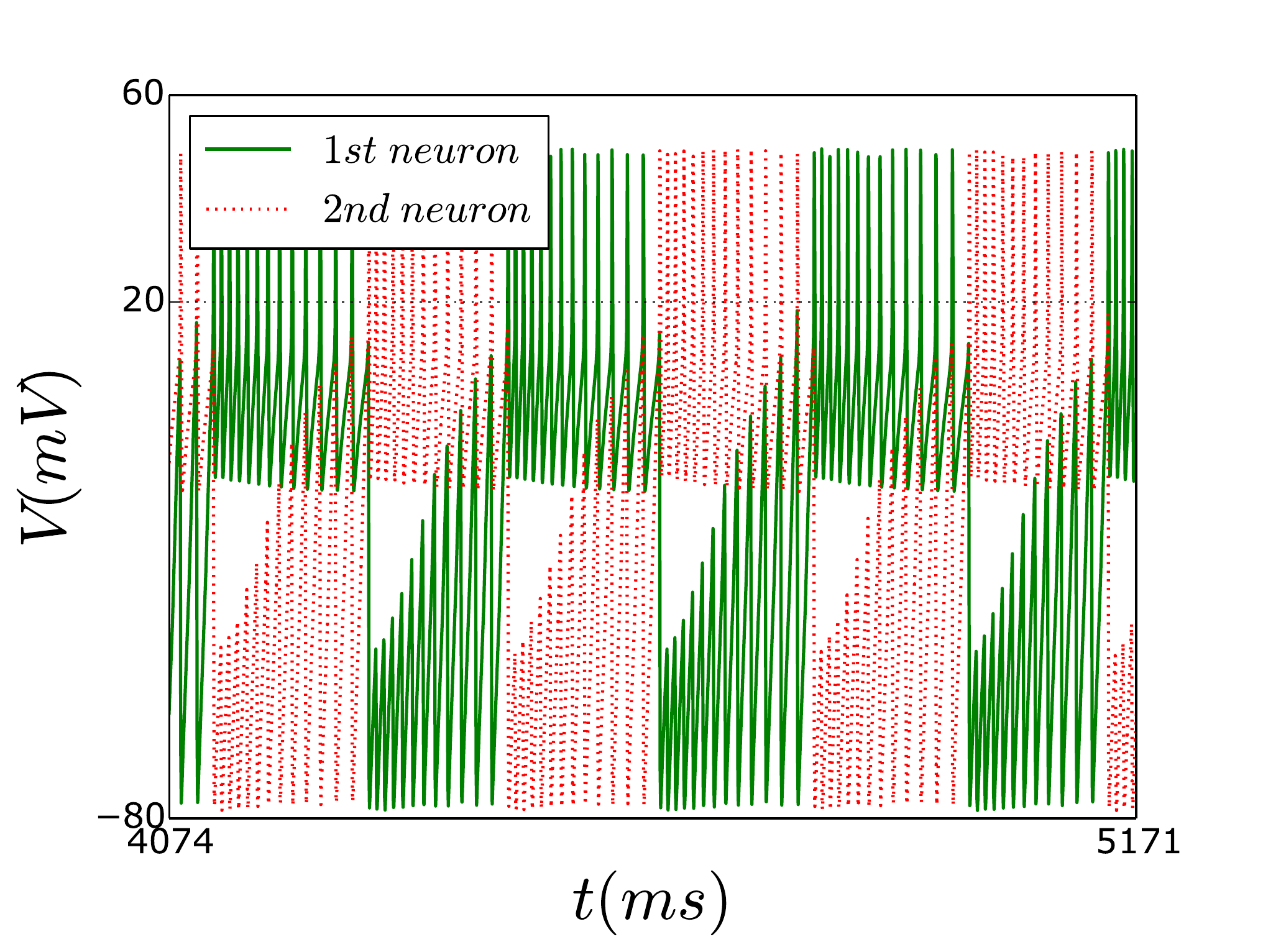}}} 
\caption{Voltage oscillatory  activity generated by the HCO at $ g_{Ca} = 1\,mS/cm^{2}$ and $I_{ext\:1}=5 $, and $\Delta I_{ext}=0.02\,\mu A /cm^{2}$:
(a) antiphase spiking at $G=0.2\,mS/cm^{2}$;  (b) chaotic spiking activity at $G=0.3\,mS/cm^{2}$;  (c) forced sub-threshold oscillations in neuron 2  due to fast spiking in neuron 1 at $G=2\,mS/cm^{2}$, (d)  network bursting at $G=4\,mS/cm^{2}$.}\label{fig:tonic_case_allG}
\end{figure}

In this section we study  emergence of alternating bursting in a HCO made of two coupled neurons that are tonically spiking in isolation at $I_{ext}^{(1)}=5\,\mu A /cm^{2}$ with relatively small $\Delta I_{ext}=0.02\,\mu A /cm^{2}$.  Our goal is to reveal how increasing the inhibitory coupling strength transforms such tonic spikers into network bursters, as illustrated in Fig.~\ref{fig:tonic_case_allG}. 

Weak coupling gives rise to periodic antiphase tonic spiking in the network  at $G=0.2\,mS/cm^{2}$ (Fig.~\ref{fig:tonic_case_allG}(a)). 
With further increases of the coupling strength, the HCO neurons evolve through a window of irregular, unreliable spiking at $G=0.3\,mS/cm^{2}$, towards an asymmetric network dynamics. In this robust regime, either neuron generates tonic spiking activity, while the other one is forced to produce  sub-threshold oscillations of quite a large amplitude, such as ones shown  in Figs.~\ref{fig:tonic_case_allG}(b,c) at at $G=2\,mS/cm^{2}$. A dramatic increase in the inhibitory coupling strength beyond $G\approx 3\,mS/cm^{2}$ finally forces the neurons to begin bursting in alternation. 
 
\begin{figure}[!ht]
\centering
\subfigure [][] {\resizebox*{.45\columnwidth}{!}{\includegraphics{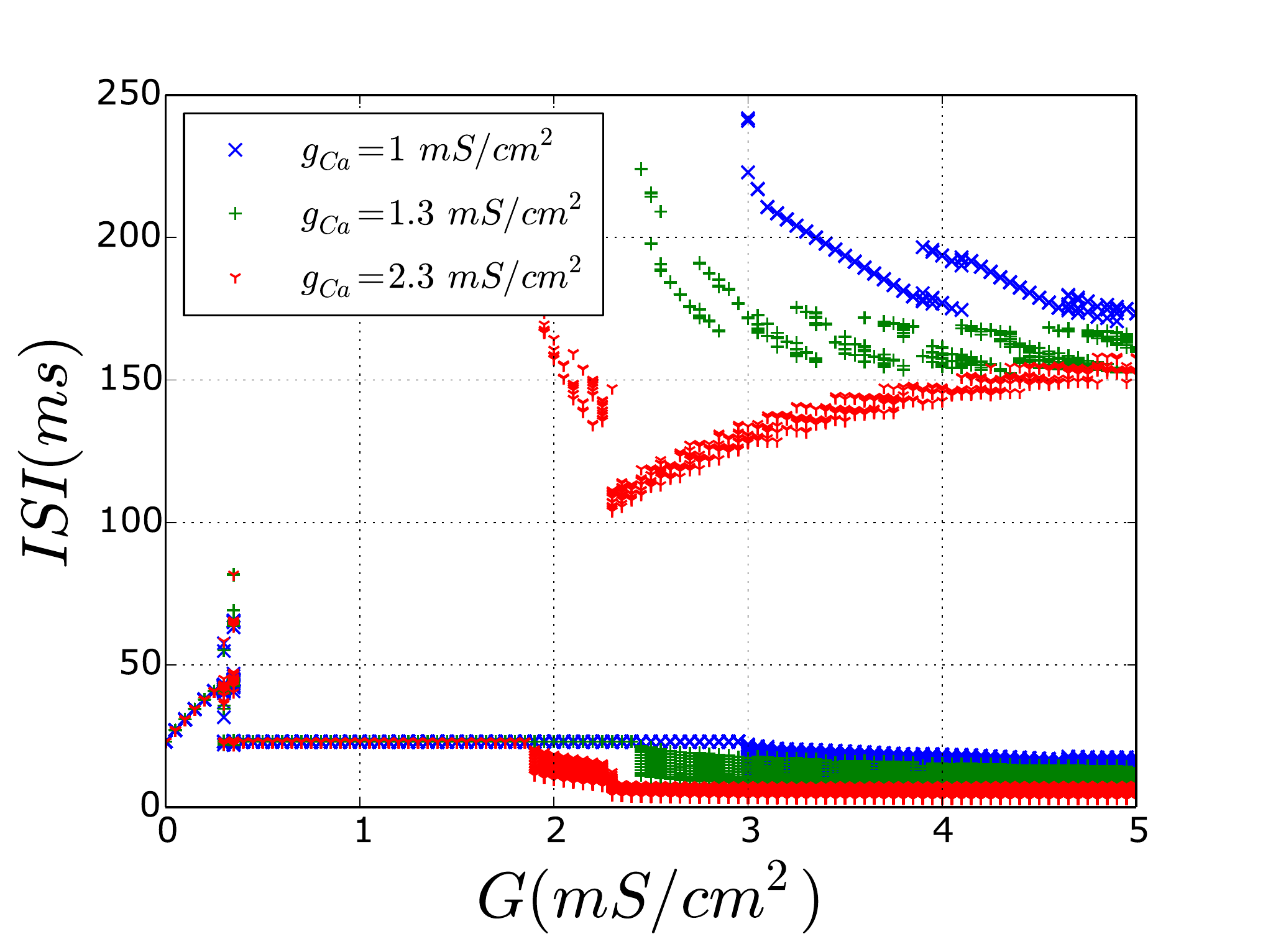}}} 
\subfigure [][] {\resizebox*{.45\columnwidth}{!}{\includegraphics{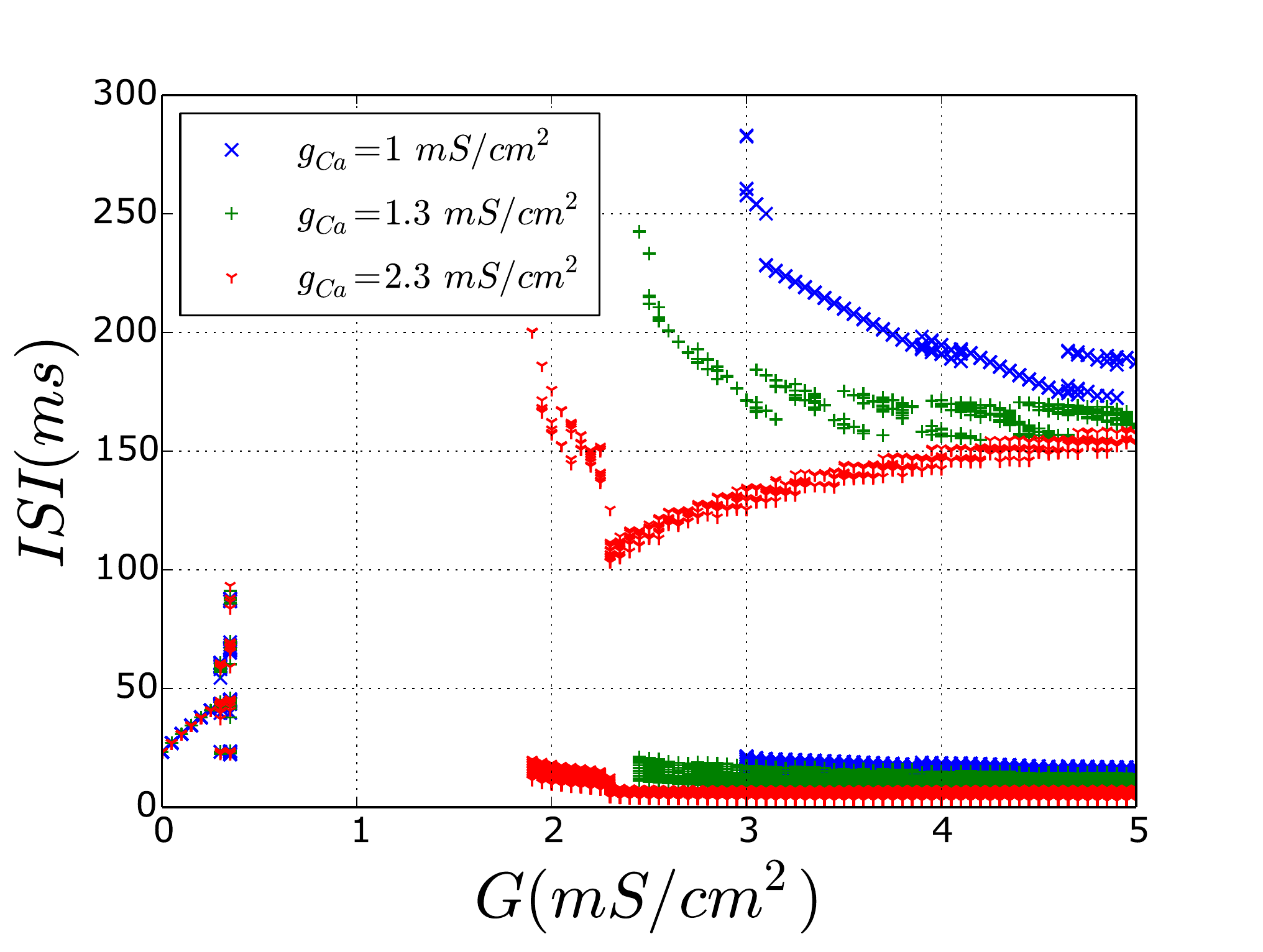}}} 
\caption{Bifurcation diagram of inter-spike intervals for coupled neuron 1 (a) and neuron 2 (b) plotted against the coupling constant $G$  for three increasing values of $g_{Ca}$; here $I_{ext\:1}=5\,\mu A /cm^{2} $ and $\Delta I_{ext}=0.02\,\mu A /cm^{2} $.}
	\label{fig:tonic_sp_ISI}
\end{figure}

Such bursting activity is often referred to as a network bursting, which is the result of strongly reciprocal interactions of two individually tonic spiking neurons. Figure~\ref{fig:IsolatedNeuronRegimes} provides an explanation why inhibitory coupling must be so strong to induce network bursting in the given model. There is a wide gap between the $I_{ext}$ parameter values corresponding to bursting (at the low end)  and to tonic spiking (on the high end). Within this gap, the neurons remain hyperpolarized quiescent, and hence  the network bursting can only occur through PIR mechanisms. A strong flux of inhibitory current is required to originate from the presynaptic, tonic spiking neuron in order to make the postsynaptic neuron drive over the gap to become a network burster. Then, the neurons  swap the roles of the driving and driven neurons and so forth.           
This assertion is further supported by Fig.~\ref{fig:tonic_sp_ISI} plotting inter-spike intervals (ISIs) in the driven postsynaptic neurons (neurons 1 and 2 in Insets (a) and  (b), resp.) against the coupling strength for three values of $g_{Ca}$.
We can see the corresponding gap in the right panel, within which neuron~2, through strongly inhibitory force, generates large sub-threshold oscillations, partially because of the network asymmetry due to $\Delta I_{ext}=0.02\,\mu A /cm^{2}$. 

Figure~\ref{fig:tonic_sp_ISI} also discloses the qualitative role of the slow low-threshold $ Ca^{2+} $-current $ I_{T}$; namely,    increasing the corresponding rebound parameter $g_{Ca}$ lowers the threshold of coupling-induced bursts, and shrinks the parameter interval of hyperpolarized quiescence in the neurons.   

\section{PIR mechanism for antiphase bursting}\label{sec:2en}

\begin{figure}[!ht]
	\centering
\subfigure [][] {\resizebox*{.45\columnwidth}{!}{\includegraphics{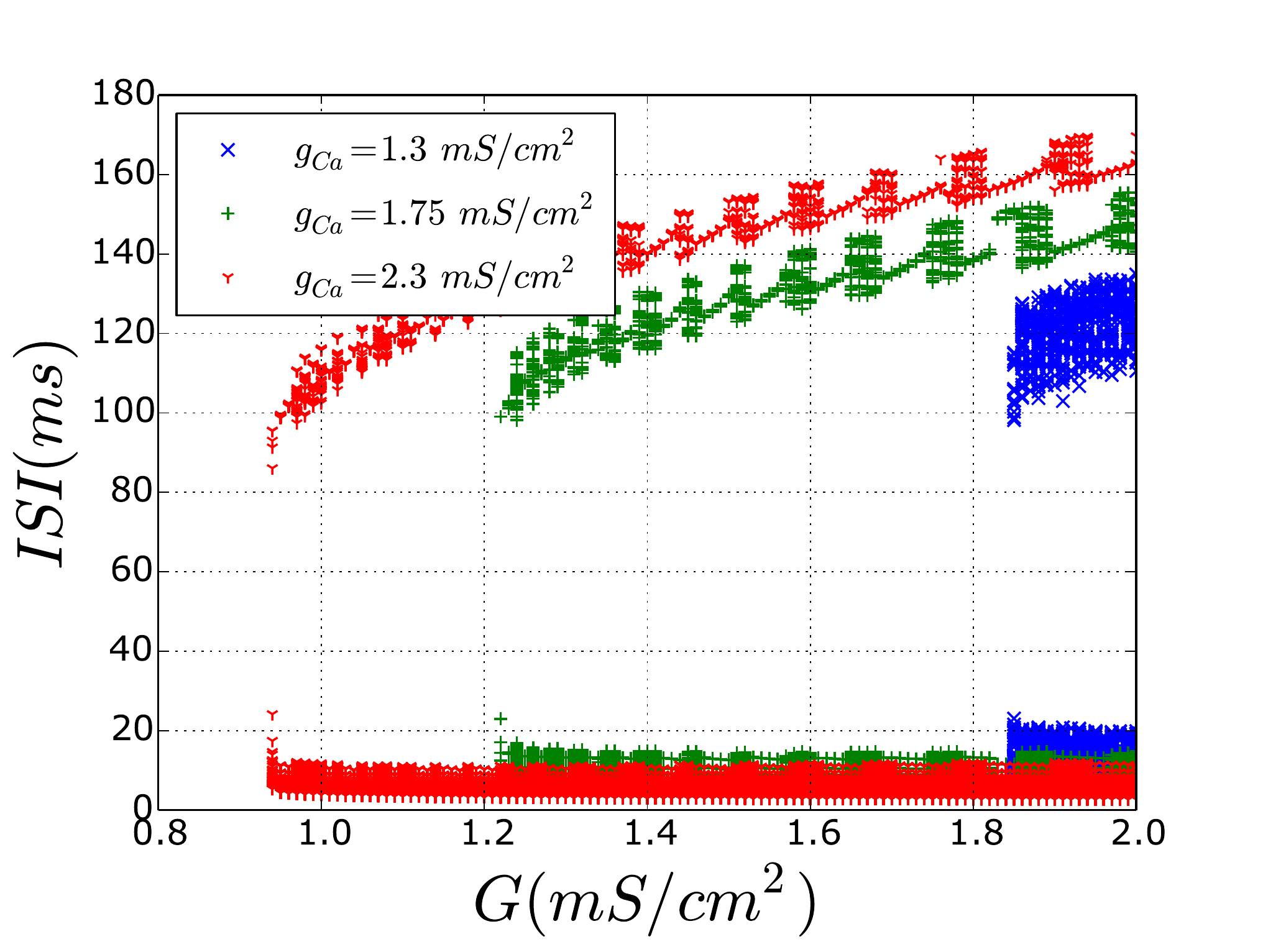}}} 
\subfigure [][] {\resizebox*{.45\columnwidth}{!}{\includegraphics{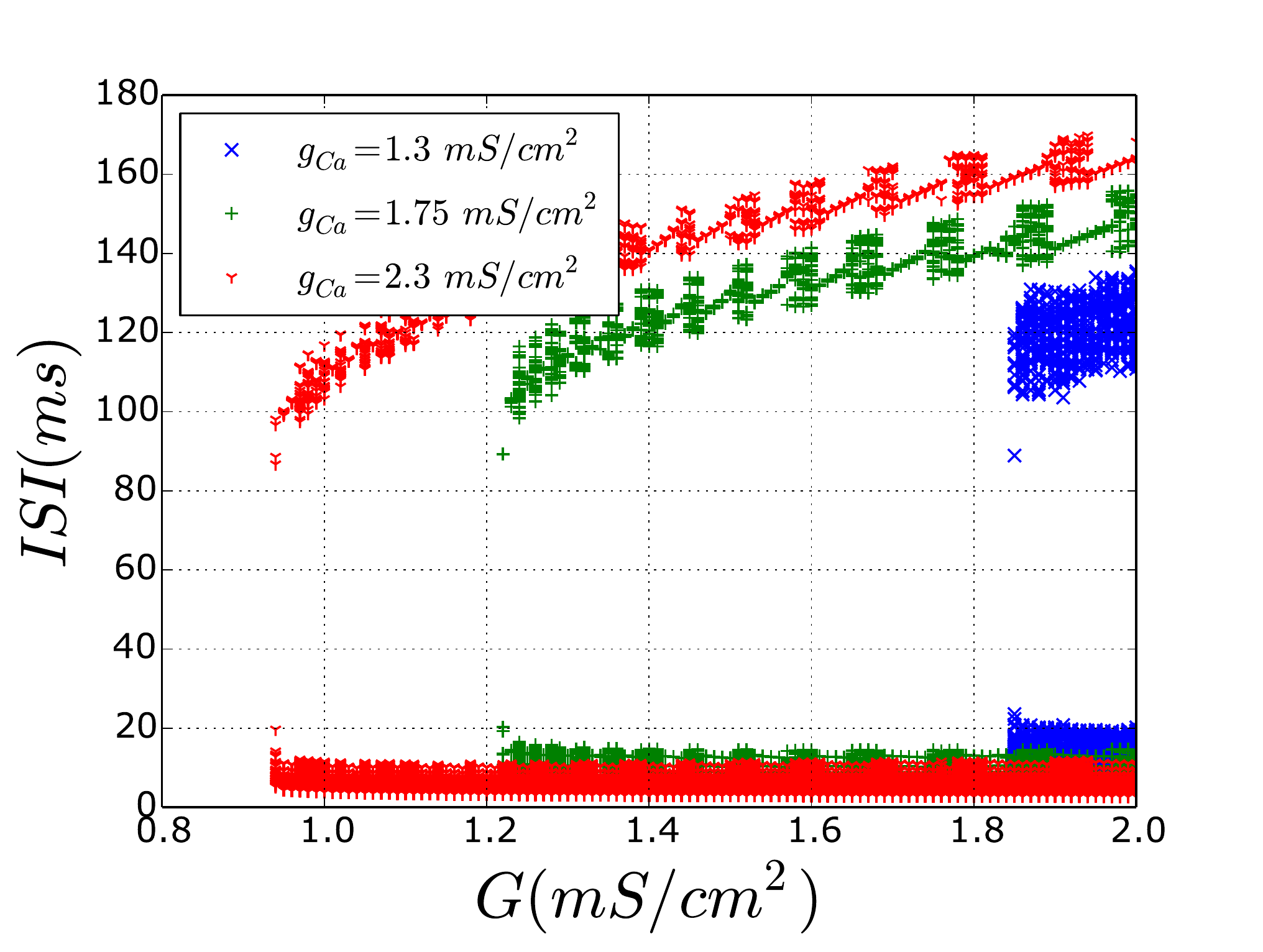}}} 
	\caption{Bifurcation diagram of ISIs evolutions for coupled neuron 1 (a) and neuron 2 (b) plotted against the coupling constant $G$; here $I_{ext}^{(1)}=2\,\mu A /cm^{2}$ and 	$\Delta I_{ext}=0.02\,\mu A /cm^{2}$ correspond to hyperpolarized quiescent neurons in isolation.}
	\label{fig:ISI_excit}
\end{figure}

In this section we examine the emergence of network antiphase bursting through the PIR mechanism.  Here, it is imperative that both neurons remain hyperpolarized within a parameter window 
quiescence $0.5\,\mu A /cm^{2}\lesssim I_{ext} 	\lesssim 4\,\mu A /cm^{2}$, as seen from Fig.~\ref{fig:IsolatedNeuronRegimes}. It is also necessary  
for PIR network bursting that the initial states of the neurons must be different: one tonic spiking and one hyperpolarized  
quiescent. An alternative is an application of negative pulse of current to trigger a PIR in a targeted neuron.  
In addition to the above constrains, the coupling strength must exceed a certain threshold, see Fig.~\ref{fig:ISI_excit}.     

It is seen from this plot that relatively  weak coupling cannot initiate PIR network bursting. Rather some sub-threshold oscillations
are generated in the post-synaptic neuron, which  cease as soon as the pre-synaptic neuron ends its active spiking phase and becomes hyper-polarized quiescent as well. For each set of the parameters there is a threshold for the coupling strength beyond which the HCO robustly produces anti-phase bursting. Figure~\ref{fig:ISI_excit} also demonstrates that the PIR mechanism of network bursting becomes more reliable with increasing $g_{Ca}$, which lowers the threshold value of inhibitory coupling.  

\begin{figure}[!ht]
	\centering
\subfigure [][] {\resizebox*{.45\columnwidth}{!}{\includegraphics{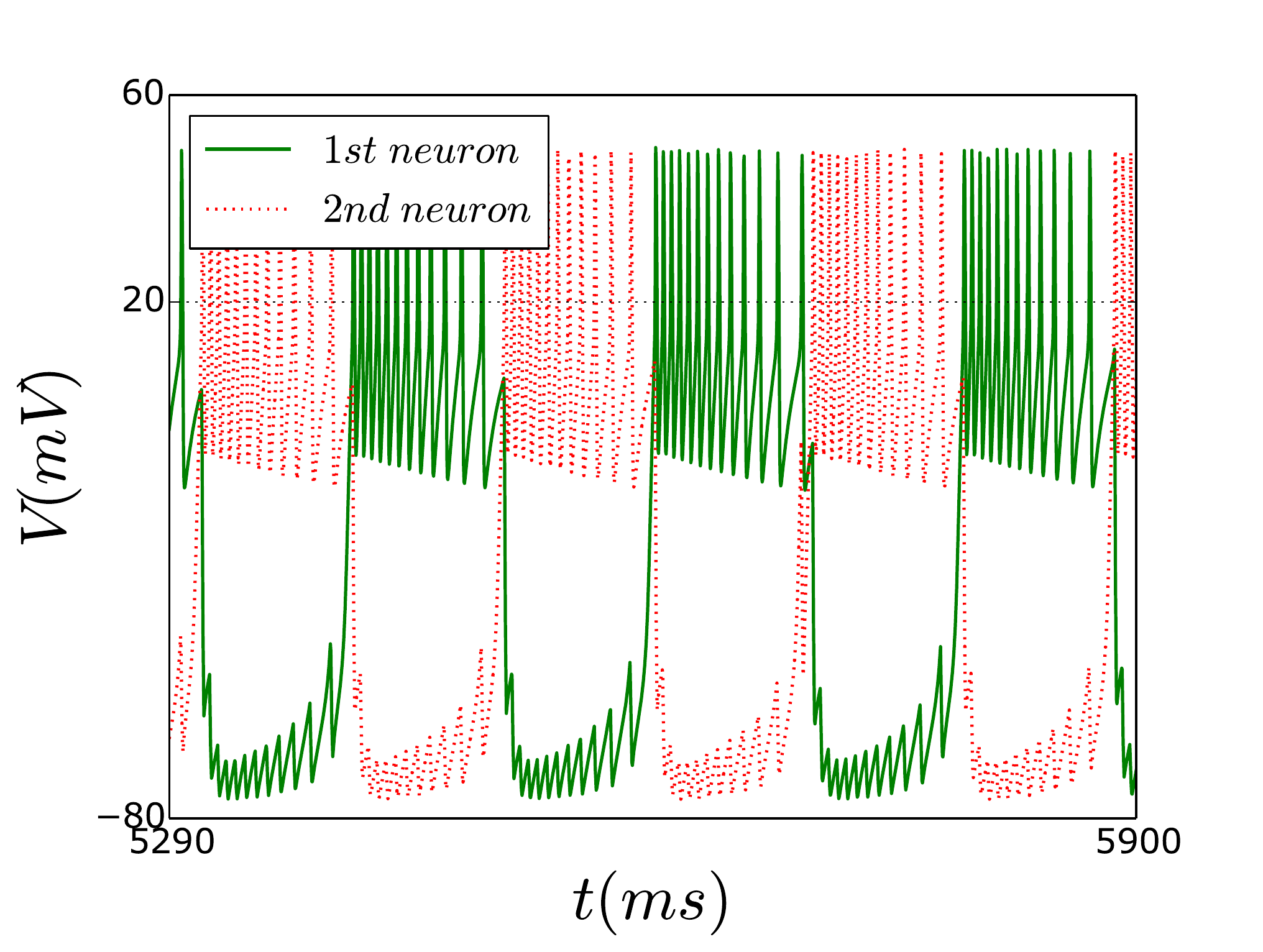}}} 
\subfigure [][] {\resizebox*{.45\columnwidth}{!}{\includegraphics{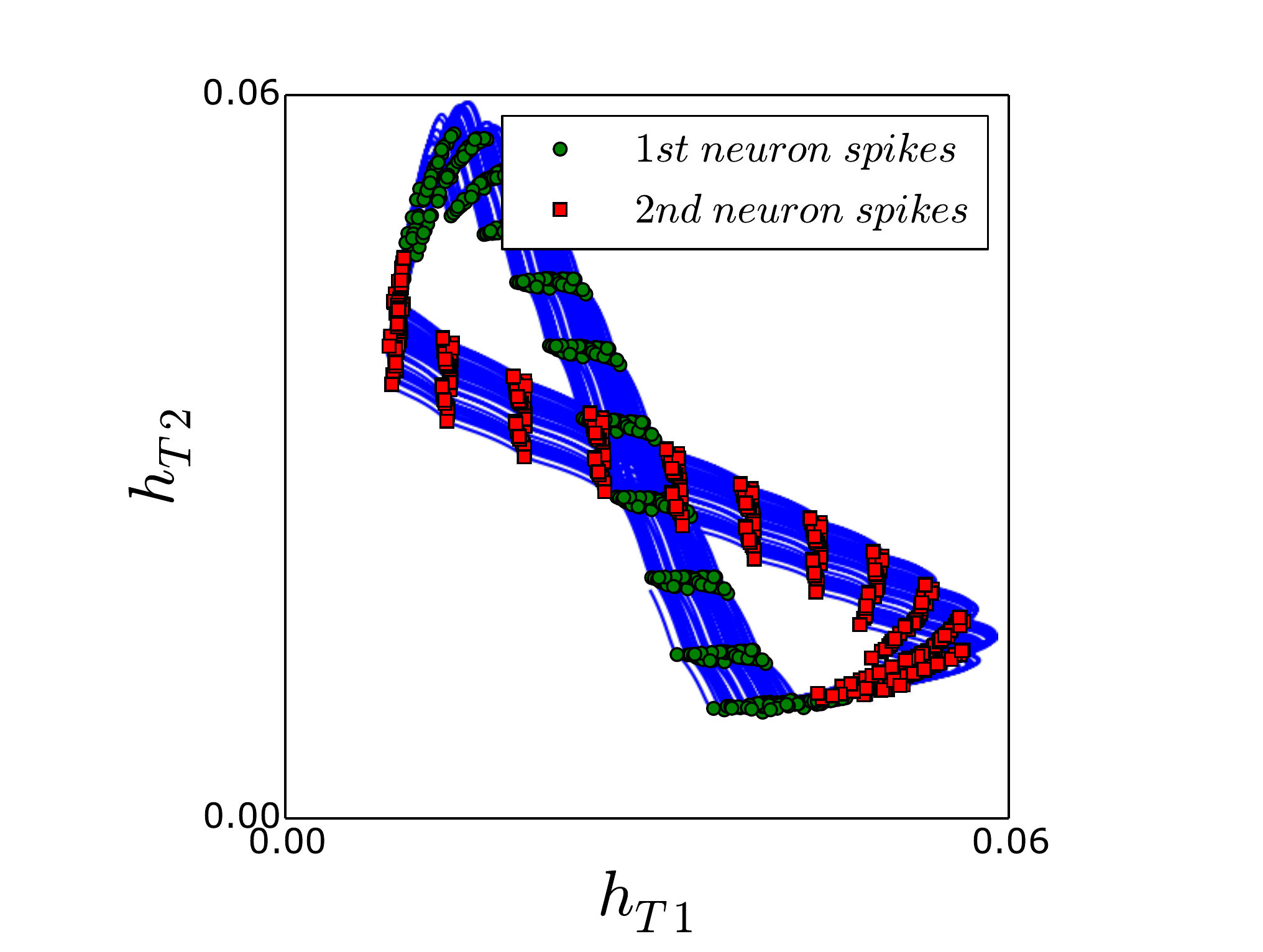}}} 
\caption{(a) Antiphase network bursting through the PIR mechanism is filled out densely by alternating voltage traces  without
quiescent gaps unlike the case of coupled endogenous bursters through the release mechanism. (b) Lissajous curve of a figure-eight shape
traced down by the slow gating variables of the models at $ G = 1.24\,mS/cm^{2} $ and $ g_{Ca} = 1.75\,mS/cm^{2}$; other parameters the same as in Fig.~\ref{fig:ISI_excit}). Circles and squares in panel (b) mark occurrences of spikes.}
	\label{fig:excit_case_G_1_24}
\end{figure}

Regularization of PIR network bursting is quite sensitive to variations of the coupling strength. By inspecting the top ISI-branch in Fig.~\ref{fig:ISI_excit} one can identify the stability windows of $G$-values corresponding to plateau-like ISI  intervals alternating 
with the windows of fluctuations of ISI values. Fluctuations are not marginal -- about  10\% off the mean value. The PIR network bursting is illustrated in Fig.~\ref{fig:excit_case_G_1_24}(a). Fig.~\ref{fig:excit_case_G_1_24}(b) reveals a quasi-periodic modulation of the slow gating variables; here the periodic Lissajous curve has a shape of figure-eight.

\section{Coupling of neurons in different modes}
\label{sec:2dm}

\begin{figure}[!ht]
\centering
\includegraphics[width=0.49\columnwidth]{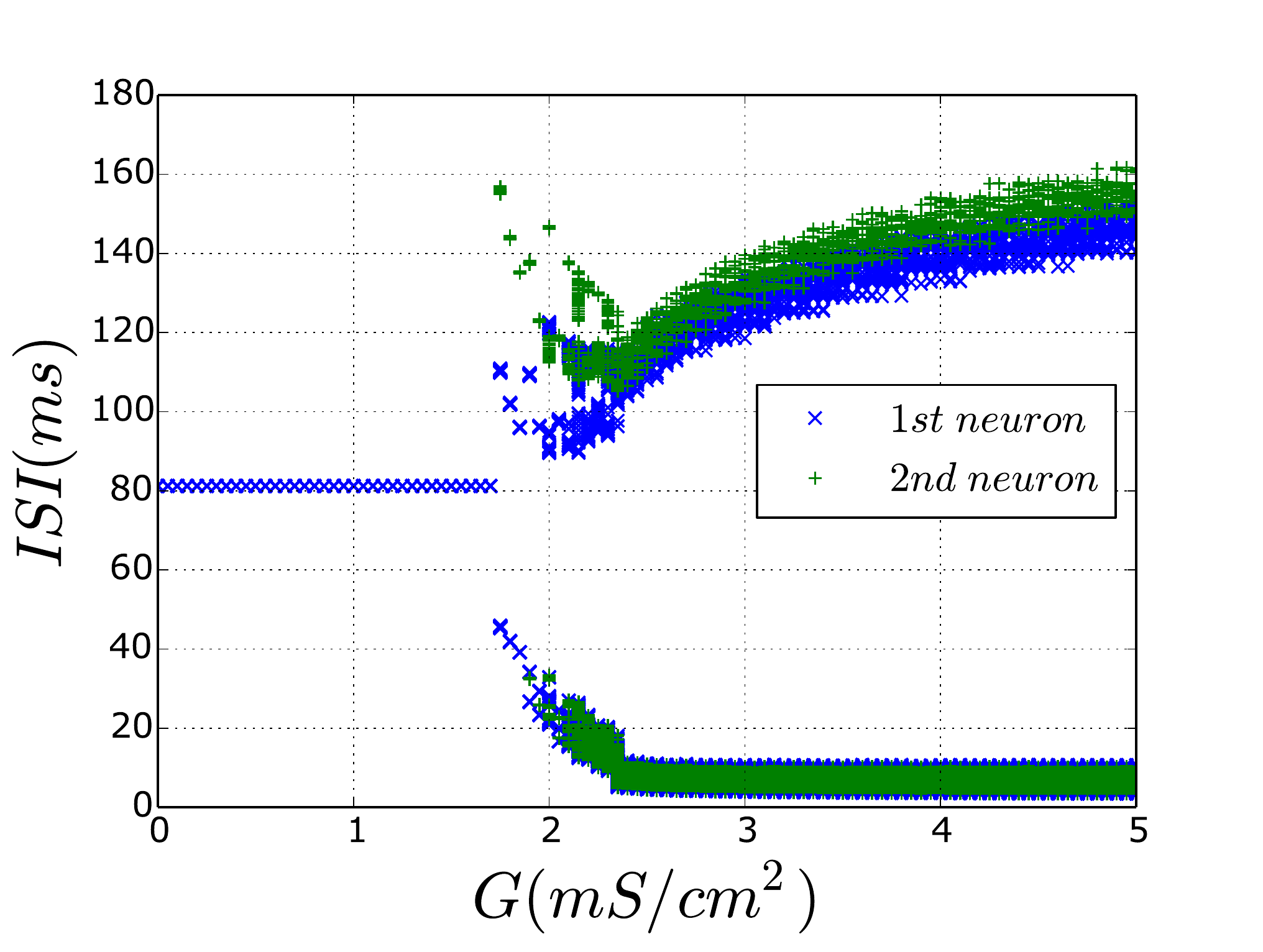}
\caption{Bifurcation diagram of ISI plotted against the coupling parameter $G$ indicates the threshold beyond which the network made of the tonically spiking neuron 1 and the hyperpolarized quiescent neuron 2 begins generating anti-phase bursting; $ g_{Ca} = 1.75\,mS/cm^{2}$.}
\label{fig:ISI_mixed_tonic_excit}
\end{figure}

\begin{figure}[!ht]
\centering
\subfigure [][] {\resizebox*{.32\columnwidth}{!}{\includegraphics{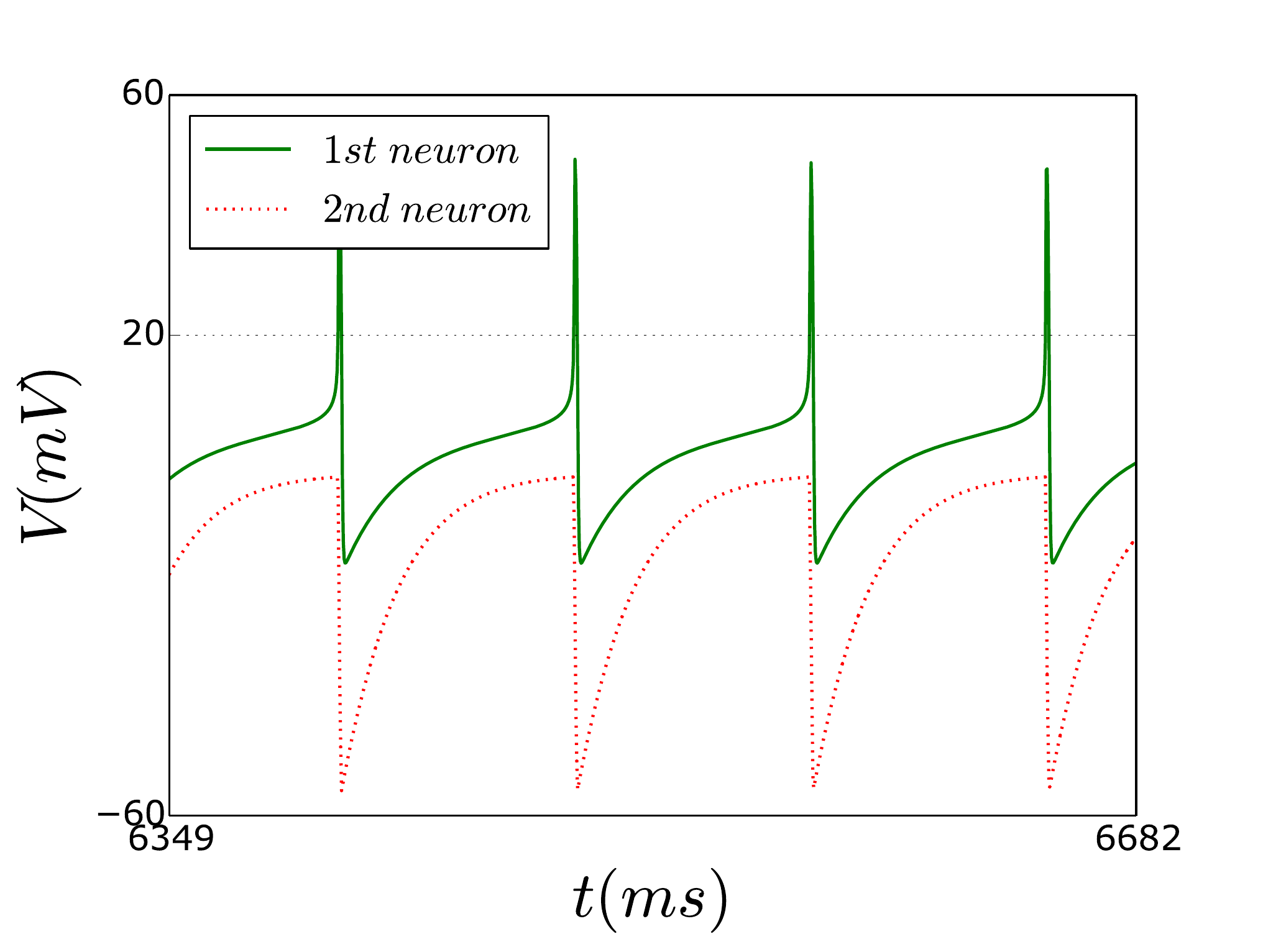}}} 
\subfigure [][] {\resizebox*{.32\columnwidth}{!}{\includegraphics{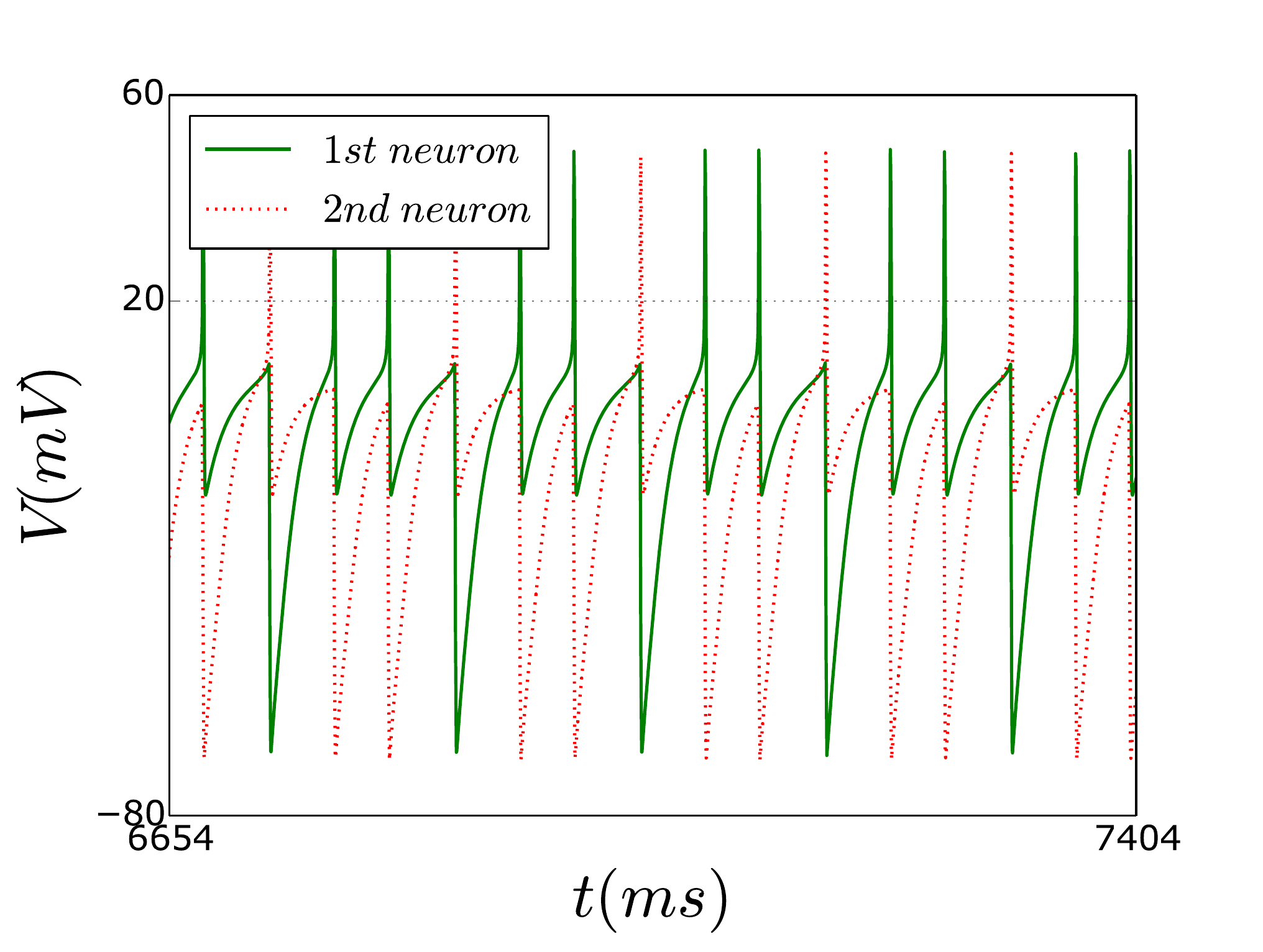}}} 
\subfigure [][] {\resizebox*{.32\columnwidth}{!}{\includegraphics{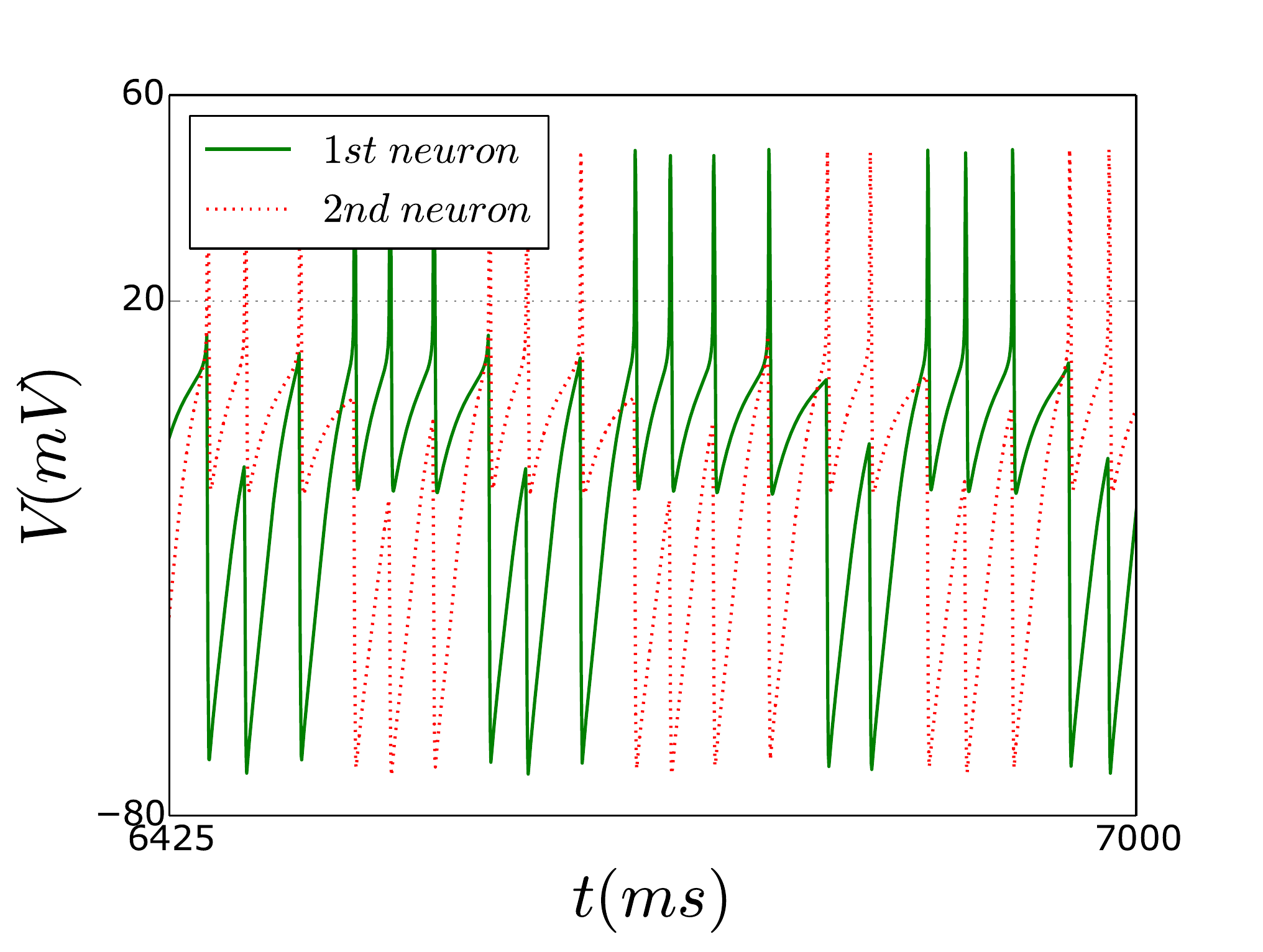}}} 
\caption{Established voltages traces generated by the HCO neurons: (a) tonic spiking in neuron 1 and inhibition induced subthreshold oscillations in neuron 2 at $G=1\,mS/cm^{2}$; (b) periodic antiphase spiking with a 2:1 locked ratio  at $G = 1.8\,mS/cm^{2}$; (c) chaotic antiphase bursting near the threshold at $G=2\,mS/cm^{2}$.}
\label{fig:mixed_tonic_excit_allG}
\end{figure}

In this section we consider the network dynamics of coupled neurons operating in different modes. Specifically, the neurons are set close to  ``quiescence-spiking'' and ``quiescence-bursting''  transitions,  respectively. 
As Fig.~\ref{fig:IsolatedNeuronRegimes} suggests, we choose the following values of the external drive: $I_{ext}^{(1)}=4.1\,\mu A /cm^{2}$ and 
$I_{ext}^{(2)}=3.7\,\mu A /cm^{2}$. This means that in isolation neuron 1 produces tonic spiking activity, while neuron 2 is hyperpolarized and ready for the PIR mechanism.  Our network simulations of two such neurons are summarized in Figs.~\ref{fig:ISI_mixed_tonic_excit} and \ref{fig:mixed_tonic_excit_allG}. We can conclude that while coupling remains weak, neuron 1 tonically spikes (Fig.~\ref{fig:mixed_tonic_excit_allG}(a)). Above a threshold, $G\approx 1.75\,mS/cm^{2}$, 
neuron 2 begins spiking as well. Fig.~\ref{fig:mixed_tonic_excit_allG}(b)) depicts alternating spiking  at $G=1.8\,mS/cm^{2}$ with 
the locking ratio 2:1, which means two spikes in neuron 1 are followed by a single spike in neuron 2.  The network begins to burst in antiphase, first with irregularly varying spike numbers, as $G$ exceeds $2\,mS/cm^{2}$, as illustrated in Fig.~\ref{fig:mixed_tonic_excit_allG}(c). Further increase in the coupling strength above $G=2.5\,mS/cm^{2}$, regularizes network bursting that becomes stable and similar to the pattern shown in Fig.~\ref{fig:excit_case_G_1_24}. 

\begin{figure}[!ht]
\centering
\includegraphics[width=0.49\columnwidth]{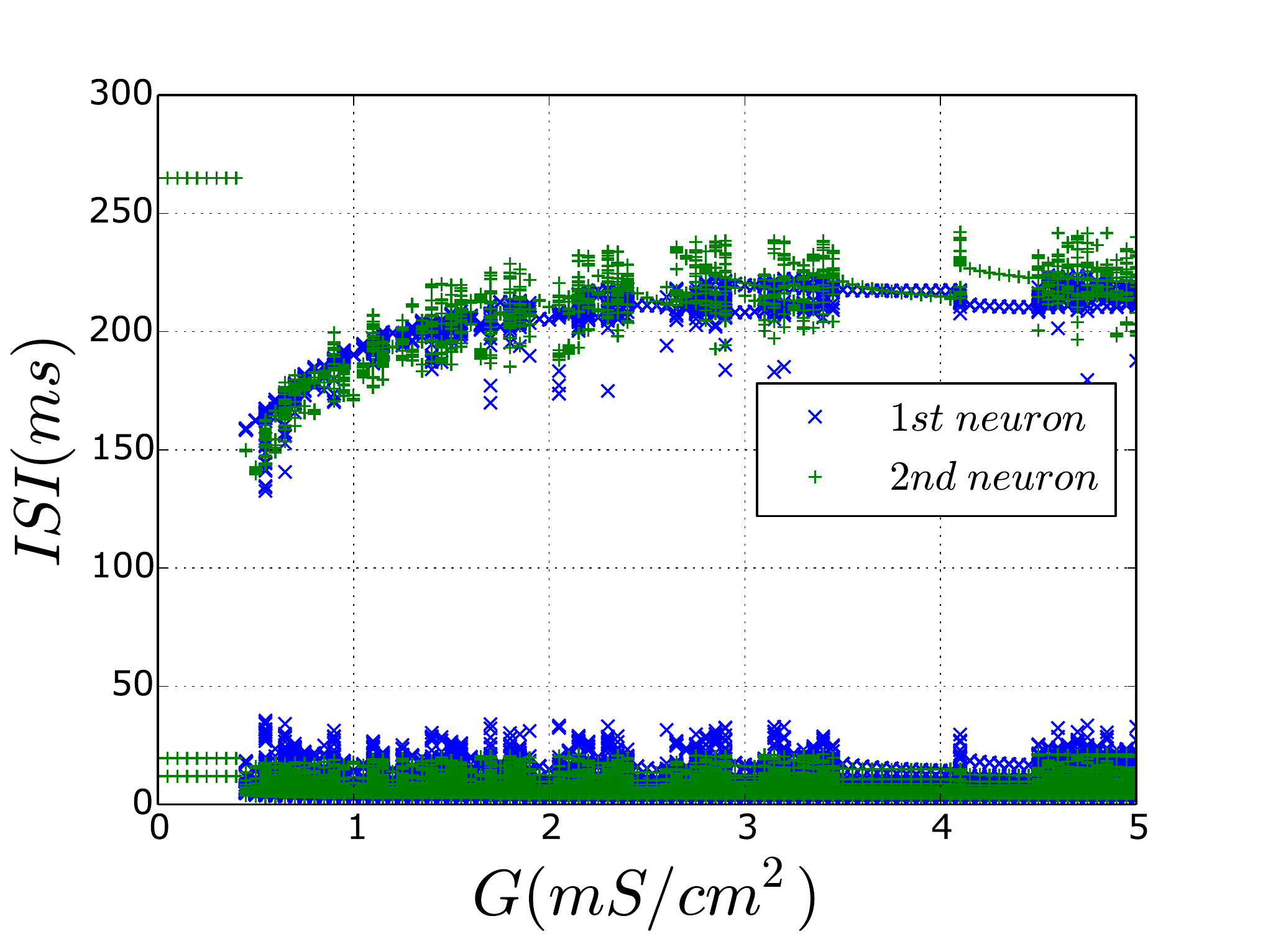}
\caption{
Bifurcation diagram of ISI plotted against the coupling parameter $G$ indicates the threshold beyond which the network made of the bursting neuron 2 and the hyperpolarized quiescent neuron 1 begins generating anti-phase bursting; $ g_{Ca} = 1.75\,mS/cm^{2}$.}
	\label{fig:ISI_mixed_burst_excit}
\end{figure}

Figure~\ref{fig:ISI_mixed_burst_excit} represents the bifurcation diagram for the network of an endogenous burster --  neuron 2 at $I_{ext}^{(2)}=0.4\,\mu A /cm^{2}$, coupled with neuron 1 -- hyperpolarized at $I_{ext}^{(1)}=1\,\mu A /cm^{2}$. Because the quiescent neuron 1 initially remains below the synaptic threshold,  anti-phase bursting in the network may only start when the PIR mechanism is induced by the endogenous burster. This explains a qualitative resemblance of the diagram in Fig.~\ref{fig:ISI_mixed_burst_excit}  with that in Fig.~\ref{fig:ISI_mixed_tonic_excit} for the HCOP made of tonic-spiking and quiescent neurons.  

\section{Conclusions}

This study focused on the mechanisms of rhythmogenesis of anti-phase bursting in HCO networks consisting 
of two reciprocally inhibitory coupled neurons. Such HCOs are primary building blocks for larger neural networks, including CPGs controlling a plethora of locomotion behaviors in spineless animals and mammals.  
There is a growing consensus in the neuroscience community that CPGs share some universal principles of their functioning.    
  
In this study, for CPG neurons we used the biophysically plausible Hodgkin-Huxley-type model. Its main feature is post-inhibitory rebound dynamics: a quiescent neuron is able to produce a single or a series of spikes after it has been quickly released from inhibition  by another pre-synaptic neuron, or by a hyperpolarized pulse of external current. The PIR mechanism allows a pair of such neurons to stably generate anti-phase bursting in a network initiated externally by inhibitory perturbation(s).

We considered several configurations of  HCOs including coupled endogenous bursters. We also discussed HCOs involving  tonic spiking and quiescent neurons that become network bursters only when coupled by fast inhibitory synapses. 
In our examination of synchronization properties of bursting we  found that, in all considered cases, the network can reliable achieve synchrony in anti-phase bursting. We also described some partial configurations leading to incomplete  synchronization where the neurons become partially synchronized, through slow-varying currents. Meanwhile, cross-correlations in their fast voltage dynamics are not always obvious. These fast voltage dynamics may give rise to the emergence of synergistically complex states, including chaos in the neural network. 

We found that while enhancing the PIR mechanism does not lead to 
drastic changes in the  dynamics of the individual neurons, it causes significant modulation of network dynamics. Specifically, we found that the windows of anti-phase bursting rhythms can be extended in the parameter space of the network, when increasing the PIR mechanisms in individual neurons become more prominent. This suggests that the PIR is a key component for robust and stable anti-phase bursting in HCOs.  In the future, we plan to examine specific networks consisting of several HCOs, which have been identified in swim CPGs of several sea mollusks.

\acknowledgments
The research is supported  by the grant (the agreement of
August 27, 2013 N 02.B.49.21.0003 between The Ministry of education and
science of the Russian Federation and Lobachevsky State University of Nizhni Novgorod, Sections 2-4)
and by the Russian Science Foundation (Project No. 14-12-00811, Sections 5-7). 
A.P. thanks A. Politi for interesting discussions and acknowledges the G.Galilei Institute 
for Theoretical Physics (Italy) for the hospitality and the INFN for partial 
support during the completion of this work. 
M.K. thanks Alexander von Humboldt foundation for support. A.S. also acknowledges the support from NSF grants DMS-1009591, RFFI 11-01-00001, and RSF grant 14-41-00044. We thank A. Kelley for helpful suggestions.

\section{Appendix: Conductance based model}
\label{appendix}


The model in this study is adopted from Ref.~\cite{DESTEXHE-CONTRERAS-SEJNOWSKI-STERIADE-94}. The dynamics of the membrane potential, $V$ is governed by
the following equation:
\begin{equation} \label{eq:v_dot}
C_{m}  V'=I_{ext} -I_{T}-I_{leak}-I_{Na}-I_{K}-I_{K[Ca]}-I_{syn};
\end{equation}
%
here, $ C_{m} $ is the specific membrane capacity, 
$ I_{ext} $ is the external current in $nA$, $ I_{T} $ is the slow low-threshold $ Ca^{2+} $-current, 
$ I_{leak} $ is the leakage current, $ I_{Na} $ is the $ Na^{+} $-current, 
$ I_{K} $ is the $ K^{+} $-current, $ I_{K[Ca]} $ is the slow $ Ca^{2+} $-activated $ K^{+} $-current
which moderates  bursting activity in the model, and $ I_{syn} $ is the synaptic current from other neurons.
Leak current $ I_{leak} $ is given by
\begin{equation} \label{eq:I_leak}
I_{leak} = g_{Na} (V - E_{leak})
\end{equation}
with $ E_{leak} = -78 \: mV $ being reversal potential for leak current,  and maximal conductance 
$ g_{L} = 0.05 \: mS/cm^{2} $.
Dynamics of fast $ Na^{+} $-current $I_{Na} = g_{Na} m^{3}  h (V - E_{Na})$ is described by the following equations:
\begin{equation} \label{eq:I_Na}
\begin{aligned}
m'& = \frac{0.32 (13 - V)}{e^{0.25 (13 - V)} - 1} (1 - m) - \frac{0.28 (V - 40)}{e^{0.2 (V - 40)} - 1} m\\
h'&  = 0.128 \cdot e^{\frac{17 - V}{18}} (1 - h) - \frac{4}{e^{-0.2 (V - 40)} + 1} h,
\end {aligned}
\end{equation}
where $ g_{Na}  = 100 \: mS/cm^{2} $ is the maximal conductance of $ Na^{+} $-current, 
$ E_{Na} = 50 \: mV $ is the Nernst equation potential for $ Na^{+} $-current, 
$ m $ and $ h $ are the gating variables describing  activation and inactivation of the current. 
Dynamics of fast $ K^{+} $-current $ I_{K} $ is described by 
\begin{equation} \label{eq:I_K}
\begin{aligned}
I_{K} &= g_{K} n^{4}  (V - E_{K}), \\
 n' &= \frac{0.032 (15 - V)}{e^{0.2 (15 - V)} - 1} (1 - n) - 0.5 e^{\frac{(10 - V)}{40}}  n
\end{aligned}
\end{equation}
with $ n $ being the gating activation variable; here $ E_{K} = -95 \: mV $ and $ g_{K}  = 10 \: mS/cm^{2} $.   
Dynamics of intraneuronular concentration of calcium ions $ [ Ca]$ is described 
\begin{equation} \label{eq:Ca}
[ Ca ]' = - \frac{k I_{T}}{2 F d} - \frac{K_{T}  [Ca]}{[Ca] + K_{d}}, 
\end{equation}
where the first term is a inflow through thin membrane due low-threshold $ Ca^{2+} $-current, and the second is a contribution of $ Ca^{2+} $ ion-pump, 
$ F = 96,469 \: C/mol$ is the Faraday constant, 
$ d = 1 \,\mu m $ is the membrane thickness, 
$ k = 0.1 $ is the dimension constant for $ [ I_{T} ] = \,\mu A / cm^{2} $ and $ [Ca] $, that has dimension of millimole , 
$ K_{T} = 10^{-4} \: mM/ms^{-1} $ and $ K_{d} = 10^{-4} \: mM $ are the speed constants from the Michaelis--Menten approximation of ion-pump kinetics, they are taken the way that one can observe fast output of calcium from a neuron in millisecond time scale.\\
Dynamics of slow low-threshold $ Ca^{2+} $-current $ I_{T} $ is modeled as follows: 
\begin{equation} \label{eq:i_t}
I_{T} = g_{Ca} m_{T}^{2} h_{T\:i} (V - E_{Ca}),
\end{equation}
with $ g_{Ca} = 1.75 mS/cm^{2} $, and the  gating activation  $ m $ and inactivation  $ h $ are given by 
\begin{equation} \label{eq:m_i_t}
m' = - \frac{m - m_{\infty}(V)}{\tau_{m}(V)}, \quad 
h' = - \frac{h - h_{\infty}(V)}{\tau_{h}(V)},
\end{equation}
with 
\begin{equation} \label{eq:m_i_t2}
\begin{aligned}
m_{\infty}(V) &= \frac{1}{1 + e^{-\frac{V + 52}{7.4}}}, \\
 \tau_{m}(V) &= 0.44 + \frac{0.15}{(e^{\frac{V + 27}{10}} + e^{-\frac{V + 102}{15}})},\\
h_{\infty}(V) &=  \frac{1}{1 + e^{\frac{V + 80}{5}}}, \\
 \tau_{m}(V) &= 22.7 + \frac{0.27}{(e^{\frac{V + 48}{4}} + e^{-\frac{V + 407}{50}})}.
\end{aligned}
\end{equation}
These constants were taken at temperature $ 36^{o}C $ and extraneuronular concentration calcium concentration $ [Ca]_{0} = 2\:mM $.
Equilibrium potential $ E_{Ca} $ depending on intraneuronular concentration of $ Ca^{2+} $ are found from the Nernst equation:
\begin{equation} \label{eq:E_Ca}
E_{Ca} = \bar k \frac{R \cdot T}{2F}ln \left (\frac{[Ca]_{0}}{[Ca]} \right );
\end{equation}
here $ R = 8.31441 \: J\,K/mol^{o} $, $ T = 309.15 ^{o}K $, dimensionless constant $\bar k = 1000 $ for $ E_{Ca} $ is measured in millivolts, extraneuronular concentration of calcium ions is $ [Ca]_{0} = 2 mM $.\\
Dynamics of $ Ca^{2+} $-activated $ K^{+} $-current is determined through the following equations:
\begin{equation} \label{eq:i_k_ca}
\begin{gathered}
I_{K[Ca]} = g_{K[Ca]} m_{K[Ca]}^{2}  (V_{i} - E_{K}), \\
 \mbox{where} \quad 
m' = - \frac{m - m_{\infty}([Ca])}{\tau_{m} \left ([Ca] \right)},
\end{gathered}
\end{equation}
and 
\begin{equation} \label{eq:i_k_ca1}
\begin{aligned}
\tau_{m}([Ca]) &= \frac{1}{48[Ca]^{2} + 0.03}, \\
 m_{\infty}([Ca]) &= \frac{48[Ca]^{2}}{48[Ca]^{2} + 0.03}, 
 \end{aligned}
\end{equation}
with $ g_{K[Ca]} = 10 mS/cm^2 $, $ E_{K} = -95\:mV $  $ m $ being the gating activation variable.\\
%
Synaptic currents are models using the fast-threshold modulation  paradigm \cite{FTM}:
\begin{equation} \label{eq:i_syn_curr_app}
I_{syn}(V_{i},V_{j}) = G  S(V_{j}-\theta_{syn})  (V_{i} - E_{syn}),
\end{equation}
where $ G $ is the maximal conductance of synaptic current flowing from pre-synaptic j-th neuron into the post-synaptic i-th neuron. For inhibitory coupling we set $ E_{syn} = - 80 \: mV $; the synaptic activity function $S(V(j)) $ is given by
\begin{equation} \label{eq:S_V}
S(V) = \frac{1}{1 + e^{-100(V-\theta)}}, 
\end{equation}
with the synaptic threshold $ \theta = 20\:mV $ set in a middle of fast spikes.


%
\end{document}